\documentclass[journal]{IEEEtran}

\usepackage{cite}
\usepackage{enumitem}
\usepackage{color}
\usepackage{graphicx}
\usepackage{booktabs,array}
\usepackage{multirow}
\usepackage{graphicx}
\usepackage[breaklinks, bookmarks = false]{hyperref}
\hypersetup{
  colorlinks   = true,
  urlcolor     = blue,
  linkcolor    = red,
  citecolor   = blue
}

\def\BibTeX{{\rm B\kern-.05em{\sc i\kern-.025em b}\kern-.08em
    T\kern-.1667em\lower.7ex\hbox{E}\kern-.125emX}}

\newcommand{\R}[1]{\textcolor{black}{#1}}

\begin{document}

\title{Cloud-Fog Automation: The New Paradigm towards Autonomous Industrial Cyber-Physical Systems}

\author{Jiong Jin,~\IEEEmembership{Member,~IEEE},
        Zhibo Pang,~\IEEEmembership{Senior Member,~IEEE},
        Jonathan Kua,~\IEEEmembership{Member,~IEEE},
        Quanyan Zhu,~\IEEEmembership{Senior Member,~IEEE},
        Karl H. Johansson,~\IEEEmembership{Fellow,~IEEE},
        Nikolaj Marchenko,
        and
        Dave Cavalcanti,~\IEEEmembership{Senior Member,~IEEE}

    


\thanks{Manuscript received 30 September 2024; revised 30 January 2025;
accepted 28 February 2025. Date of publication DD MONTH 2025; date
of current version DD MONTH 2025. J.~Jin's work was supported in part by the Australian Research Council Linkage Project under Grant LP190100594. Z.~Pang's work is partly funded by Swedish Foundation for Strategic Research (SSF) through project APR20-0023 and by Swedish Innovation Agency (Vinnova) through the SweWIN center (2023-00572). Q.~Zhu's work is supported in part by the National Science Foundation (NSF) under Grant ECCS-1847056. \textit{(Corresponding authors: Jiong Jin and Zhibo Pang.)}}

\thanks{Jiong Jin is with the School of Engineering, Swinburne University of Technology, Hawthorn VIC 3122, Australia. (email: jiongjin@swin.edu.au).}

\thanks{Zhibo Pang is with the Department of Intelligent Systems, KTH Royal Institute of Technology, 10044 Stockholm, Sweden, and the Department of Automation Technology, ABB Corporate Research Sweden, 72226 V\"aster{\aa}s, Sweden. (email: zhibo@kth.se).}


\thanks{Jonathan Kua is with the School of Information Technology, Deakin University, Geelong VIC 3220, Australia. (email: jonathan.kua@deakin.edu.au).}

\thanks{Quanyan Zhu is with the Department of Electrical and Computer Engineering, Tandon School of Engineering, New York University, Brooklyn,
NY 11201 USA. (email: quanyan.zhu@nyu.edu).}

\thanks{Karl H. Johansson is with the Division of Decision and Control Systems, School of
Electrical Engineering and Computer Science, and Digital Futures, KTH Royal Institute of Technology, 10044 Stockholm, Sweden. (email: kallej@kth.se).}

\thanks{Nikolaj Marchenko is with Corporate Research, Robert Bosch GmbH, 71272 Renningen, Germany. (email: nikolaj.marchenko@bosch.com).}

\thanks{Dave Cavalcanti is with Intel Corporation, Santa Clara, CA
95054 USA. (email: dave.cavalcanti@intel.com).}
}

\markboth{IEEE Journal on Selected Areas in Communications,~Vol.~XX, No.~X, Month 2025}%
{Jin \MakeLowercase{\textit{et al.}}: Cloud-Fog Automation: The New Paradigm towards Autonomous Industrial Cyber-Physical Systems}

\maketitle

\begin{abstract}

Autonomous Industrial Cyber-Physical Systems (ICPS) represent a future vision where industrial systems achieve full autonomy, integrating physical processes seamlessly with communication, computing and control technologies while holistically embedding intelligence. \R{Cloud-Fog Automation is a new digitalized industrial automation reference architecture that has been recently proposed. This architecture is a fundamental paradigm shift from the traditional International Society of Automation (ISA)-95 model to accelerate the convergence and synergy of communication, computing, and control towards a fully autonomous ICPS.} With the deployment of new wireless technologies to enable almost-deterministic ultra-reliable low-latency communications, a joint design of optimal control and computing has become increasingly important in modern ICPS. It is also imperative that system-wide cyber-physical security are critically enforced. Despite recent advancements in the field, there are still significant research gaps and open technical challenges. Therefore, a deliberate rethink in co-designing and synergizing communications, computing, and control (which we term \textit{``3C co-design''}) is required. In this paper, we position Cloud-Fog Automation with 3C co-design as the new paradigm to realize the vision of autonomous ICPS. \R{We articulate the state-of-the-art and future directions in the field, and specifically discuss how goal-oriented communication, virtualization-empowered computing, and Quality of Service (QoS)-aware control can drive Cloud-Fog Automation towards a fully autonomous ICPS, while accounting for system-wide cyber-physical security.}

\end{abstract}


\IEEEpeerreviewmaketitle

\newpage 
\section{Introduction}
\label{introduction}


\IEEEPARstart{R}{apid} technological advancements in Industrial Internet of Things (IIoT), artificial intelligence (AI), machine learning (ML), and cloud/edge computing paradigms have driven large-scale innovations in industrial automation systems in Industry 4.0 -- spanning across many verticals, such as advanced manufacturing, intelligent transportation, smart logistics, smart grids, robotics, agriculture, healthcare, energy, urban planning, and so forth~\cite{Sisinni:2018}. Modern industrial automation systems play a major role in Industrial Cyber-Physical Systems (ICPS) where physical processes aimed to be seamlessly integrated with computational and communication technologies. For instance, ICPS transforms conventional manufacturing systems into smart environments by embedding intelligence into physical processes, where machines, robots, and human operators collaborate seamlessly. These systems are characterized by their ability to monitor, analyze, and respond to changes in the environment, which ensures higher levels of operational efficiency, flexibility, and safety. However, most of these ICPS capabilities and processes are not yet fully autonomous. Therefore, there is a growing traction in academia and industry to pursue a futuristic vision of ``Autonomous ICPS" where the entire ICPS holistically functions, operates and executes industrial processes with full autonomy, while simultaneously embedding intelligence throughout the system.

\R{Autonomous ICPS will create a new technological paradigm that forms the backbone of future industrial automation systems towards Industry 5.0~\cite{xiang2023advanced}. The fundamental architecture of most industrial automation systems today stems from the International Society of Automation (ISA)-95 model, which has largely remained unchanged for the past two decades~\cite{IEC62264}. ISA-95 is a pyramid architecture, where all communication, computing, and control elements of ICPS (such as those in smart manufacturing enterprises) are both logically and physically partitioned into five separate levels, i.e., Levels 0 to 4. Each of the levels correspond to physical machines, low level machine control, middle level supervisory control, manufacturing execution systems (MES) and enterprise resource planning (ERP) systems.}

\R{With the rapid penetration of digital technologies in ICPS, there is a concerted industrial effort towards building a more network-centric architecture, where elements in any of the five ISA-95 levels can be directly connected with elements in another other levels according to system requirements. Therefore, ISA-95 has become a logical/virtual partitioning architecture for semantic definition of functionalities instead of physical partitioning/topological layout of the industrial elements. These changes have recently led to several Cloud-based Automation architectures, such as the NAMUR Open Architecture (NOA), Open Platform Communications Unified Architecture (OPC UA), and Cloud Robotics.}

\R{Cloud-Fog Automation took this development further by proposing a novel industrial system reference architecture which significantly disrupts the conventional ISA-95 layered model by fully leveraging cloud/fog computing and AI/ML technologies to achieve the convergence of communication, computing, and control in ICPS~\cite{Cloud-Fog_Automation}. It is a network-centric architecture where cloud and fog technologies are used collaboratively to provide determinism in connectivity, networked computing, and intelligent networked control in ICPS. Cloud-Fog Automation is a fundamental paradigm shift where industrial applications can be realized independent of vendor-specific hardware and software, and increasingly being provided as a service with varying degrees of technical and infrastructure requirements for flexible deployments (e.g., Control-as-a-Service, Robots-as-a-Service, Manufacturing-as-a-Service), which translates into significant capital expenditure (CapEX) and operational expenditure (OpEX) benefits.}

\R{An important characteristic of Cloud-Fog Automation is the deployment of next-generation wireless technologies to provide greater flexibility, redundancy, and mobility.} It is imperative that these wireless technologies provide ultra-reliable low-latency communications (uRLLC) for time-critical computing and control applications, without compromising functional safety and security of ICPS. \R{New wireless technologies such as 5G New Radio (NR), Wi-Fi 6/7, wireless time-sensitive networks (TSN) are required to guarantee the availability of the overall system, and satisfy the stringent latency and reliability requirements of Level-1 and Level-2 control.} For example, when the network or cloud/fog computing infrastructure is temporarily unavailable or under-performing, the orchestration system will automatically move the time-critical tasks to closer to the data producer and consumer to mitigate the impacts.

\R{Compared to traditional Level-1 and Level-2 (in the ISA-95 architecture) controllers, Cloud-Fog Automation offers significantly-enhanced computing power at lower costs, as driven by advancements in fog and edge computing technologies~\cite{Cloud-Fog_Automation}. The proximity to the field and ``physically distributed" architecture are crucial for reducing latency, improving reliability, and addressing security and safety challenges. The ``logically centralized" and virtualized nature of Cloud-Fog Automation simplifies post-commissioning tasks such as upgrades, reconfiguration, dynamic resource allocation, load balancing, and modular task management, while streamlining the digital distribution and delivery of products and services. Its compatibility with emerging AI technologies makes Cloud-Fog Automation an ideal platform for supporting next-generation ICPS, which are beyond the capabilities of conventional automation systems. When integrated with wireless networks, Cloud-Fog Automation further reduces installation and cabling costs, while enabling greater mobility, flexibility, and finer-grained deployments of automation systems.}


\R{Although Cloud-Fog Automation is a promising architectural proposal, there are still significant research gaps in realizing the vision of a fully autonomous ICPS. Specifically, the tight coupling across communication, computing and control domains is pivotal. Therefore, a deliberate rethink in co-designing and synergizing communication, computing, and control (which we term \textit{``3C co-design"} in this paper) is required to position Cloud-Fog Automation as a new paradigm towards a fully and all-encompassing autonomous ICPS.} There are currently significant gaps in understanding the intricate interplay between communication, computing, and control in this context. To this end, we have specifically identified the three targeted research domains in 3C co-design. These research domains are deemed to be the key driving forces for realizing autonomous ICPS with Cloud-Fog Automation as the transformative paradigm: (i) goal-oriented communication; (ii) virtualization-empowered computing; and (iii) Quality of Service (QoS)-aware control.

In this paper, we start by emphasizing the pressing need for 3C co-design, and clearly articulate the key constituents that are required to implement the vision of autonomous ICPS. \R{We then provide a scoping for the state-of-the-art and latest advances in goal-oriented communication, virtualization-empowered computing and QoS-aware control; and endeavor to promote research and innovations to bridge the gap between theory and real-world applications, along with their practical implications.} This paper serves as a roadmap and vision to steer research and development efforts in driving Cloud-Fog Automation as the new paradigm towards autonomous ICPS, bringing 3C co-design to the forefront.

\begin{figure*}[htp]
	\centering \includegraphics[trim={0 0cm 0 0cm},clip,width=\linewidth]{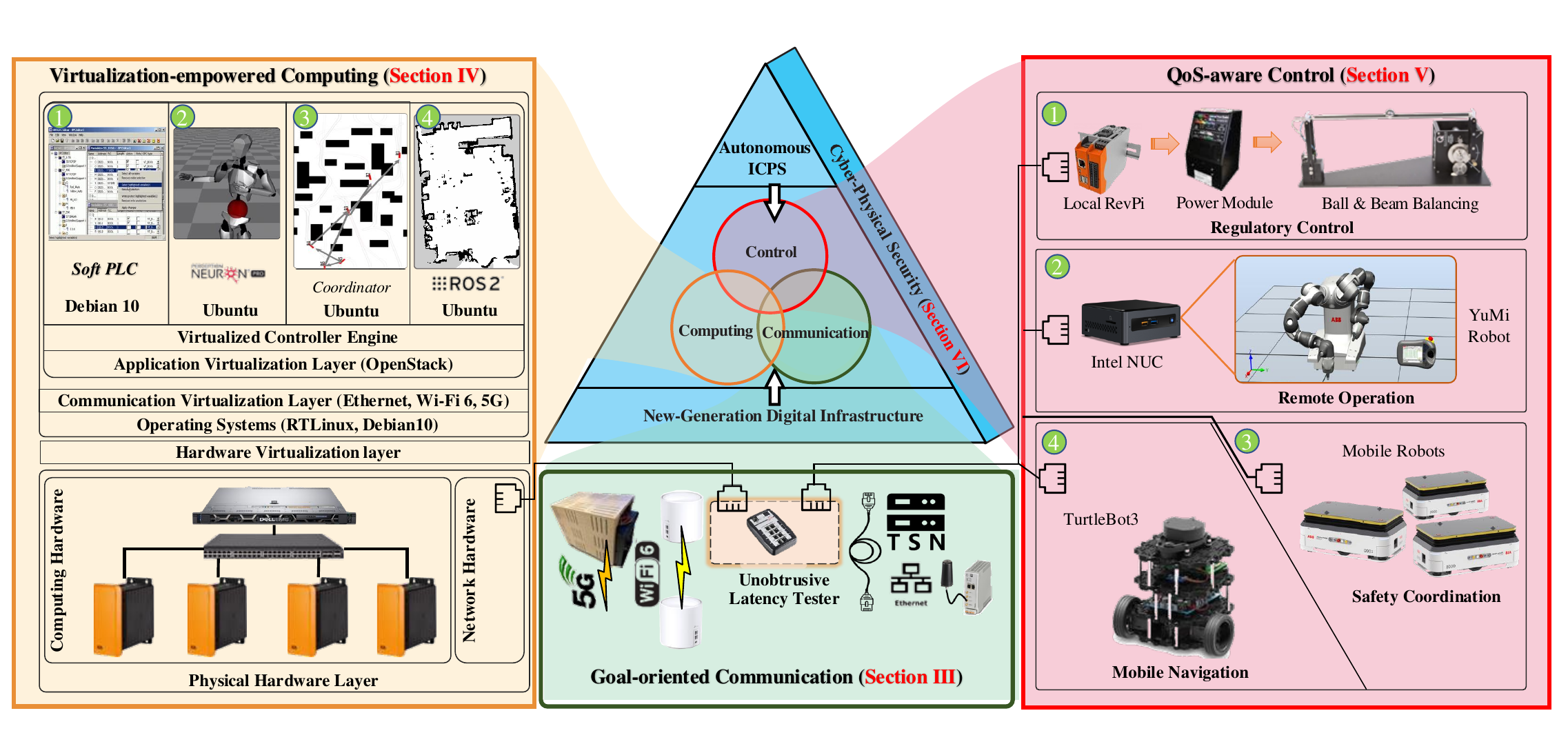}
	\caption{\R{Architectural overview of our Cloud-Fog Automation prototype with four representative use cases (adapted from~\cite{lyu10558844}): (1) regulatory control; (2) remote operation; (3) safety coordination; and (4) mobile navigation. The central pyramid figure illustrates the crucial role of 3C co-design in closing the gap between the current/emerging new-generation digital infrastructure and the vision of autonomous ICPS. Cyber-physical security cuts across the 3C domains.}} 
	\label{fig:3C-Co-design-use-cases}
\end{figure*}

The rest of the paper is organized as follows. \R{Section~\ref{section2_co-design} presents a Cloud-Fog Automation prototype with four representative use cases, outlines the design requirements of autonomous ICPS and discusses the pressing need for 3C co-design in realizing this vision. Section~\ref{section3_communication} introduces goal-oriented communication and its performance metrics, review the state-of-the-art, and outlines important research questions in rendering industrial communication systems goal-oriented. Section~\ref{section4_computing} presents the evolution of virtualization technologies, stemming from cloud computing to industrial use cases, and specifically survey virtualization techniques in ICPS across application, communication and hardware layers, and provide future research avenues at each layer. Section~\ref{section5_control} identifies the challenges of QoS-aware control in distributed control automation architectures, reviews key enabling technologies and outlines research directions. Section~\ref{section6_security} presents the critical aspect of cyber-physical security that cuts across the 3C domains, including potential security risks, attacks and defense strategies leveraging AI and large language models (LLMs) techniques.} Finally, Section~\ref{conclusions} concludes the paper.

\section{Co-Design of Communication, Computing and Control in Cloud-Fog Automation}
\label{section2_co-design}


\R{In this section, we introduce a prototype featuring four distinct industrial use cases, and outline the design requirements of autonomous ICPS. We then review technologies in current/emerging new-generation digital infrastructure, and articulate how co-designing communication, computing, and control in Cloud-Fog Automation can pragmatically close the gap to fully realize the vision of autonomous ICPS.}

\subsection{\R{Representative Use Cases through 3C Co-design in Cloud-Fog Automation Prototype}}

\R{To illustrate the significance of 3C co-design and transformative potential in ICPS, we designed and implemented a real-world Cloud-Fog Automation prototype with four representative use cases in the industrial automation system context: (1) regulatory control; (2) remote operation; (3) safety coordination; and (4) mobile navigation. These four use cases are crucial features of modern ICPS with varying reliability and latency requirements. More specially, these four use cases present critical industrial \textit{control} applications, that are enabled by the underlying \textit{communication} networks and \textit{computing} infrastructure. Fig.~\ref{fig:3C-Co-design-use-cases} presents an architectural overview of these four use cases and their respective capabilities supported by communication, computing and control technologies within the central `3C co-design pyramid' figure. We investigate the impacts of current communication and computing technologies on control applications' performance, and subsequently derive key 3C co-design insights/design requirements for future autonomous ICPS. We refer the reader to~\cite{lyu10558844} for further implementation details and result analysis.}

\R{In the first use case, the \textit{regulatory control} process ensures compliance with industrial standards by overseeing and managing the kinematic and dynamic behaviors of systems. Cloud-Fog Automation facilitates efficient monitoring and control of time-critical industrial operations, such as continuous production processes. It enables real-time system supervision and immediate feedback to maintain adherence to regulatory standards by co-designing and integrating virtualization technologies with communication frameworks. However, challenges such as unpredictable network delays, sudden interruptions, and unstable outputs from virtualized computing can lead to production line disruptions, compromised product quality, or safety risks.}


\R{The second use case focuses on \textit{remote operation}, enabling remote control of industrial equipment to protect workers from physically harsh and unsafe environments. Cloud-Fog Automation supports this by facilitating low-latency remote operation and real-time monitoring of on-site control states and signals. It achieves this through the deployment of soft controllers and data storage within the guest OS layer of the virtualized computing infrastructure. Nonetheless, unreliable communication and uncertainties arising from virtualization can lead to mismatches between operator commands and the actions executed by remote devices.}


\R{The third use case focuses on \textit{safety coordination}, which ensures compliance with safety policies in industrial environments by managing the movements of multiple robots. This requires a centralized coordinator, where robots operate based on dynamic modeling and are synchronized through specific ordering policies. Cloud-Fog Automation facilitates real-time communication among robots, enabling them to share information about the real-time positions/status of nearby objects, thus ensuring adherence to safety protocols. However, unstable communication channels may cause delays in real-time status updates or loss of control commands, potentially leading to collisions and hazardous situations.}


\R{In the fourth use case of \textit{mobile navigation}, we evaluate the accuracy and efficiency of autonomous guided vehicles (AGVs) in navigating large-scale maps. High-speed data transmission and low-latency interactions between AGVs and edge controllers is ensured through co-designing the AGVs' control requirements with communication networks with virtualization software. Cloud-Fog Automation enables the rapid and efficient processing of large data volumes, a critical requirement for robotic navigation tasks. However, the performance of robotic control in edge computing environments depends on the specific edge computing deployment strategies~\cite{Aazam2018}.}


\R{In summary, the four representative use cases presented above highlight the critical importance of 3C co-design, particularly the co-design of \textit{communication} networks and \textit{computing} infrastructure with \textit{control} requirements to support critical industrial applications. The presented prototype demonstrates that Cloud-Fog Automation has both the capability and flexibility to meet the stringent and multi-dimensional requirements of diverse industrial applications.}


\subsection{Design Requirements of Autonomous Industrial Cyber-Physical Systems} \label{subsection:Autonomous_ICPS}
\R{Although the aforementioned prototype and use cases demonstrate the significant potential of Cloud-Fog Automation, they currently lack full autonomy and require coordinated efforts to implement sophisticated autonomous industrial systems, and ultimately realize the vision of \textit{autonomous ICPS}. The flexible and open nature of the Cloud-Fog Automation architecture offers substantial opportunities not only to enhance the autonomy of ICPS but also to optimize system performance, minimize resource consumption, and ensure resilient operations across diverse application domains.} The integration of AI/ML and big data technologies within ICPS, alongside increasingly sophisticated sensing and communication capabilities, necessitates more than just timely data delivery and reliable communication. Here we outline the critical design requirements for emerging autonomous ICPS, starting with examples of key application drivers. We then explore the essential system specifications, examine how technological trends are shaping these advancements, and emphasize the importance of a holistic 3C co-design approach to effectively navigate the inherent complexities and trade-offs.

Across various industrial sectors, ICPS technologies are being extensively studied and evaluated to tackle critical challenges related to resource efficiency, real-time performance, and system resilience~\cite{anna+23}. In process automation, for example, real-time wireless sensing and networking have demonstrated significant potential to enhance flexibility and improve real-time data collection in dynamic scenarios, such as the start-up phase of a paper mill or the adjustment of a chemical process~\cite{ahlen+19}. The advent of Industry~4.0 has lowered the barriers for implementing ICPS solutions within the manufacturing industry, where Cloud-Fog Automation is poised to make early and vital contributions~\cite{daff+21}. \R{The ongoing transformation of the energy grid, characterized by increased renewable electricity production and flexible consumption, presents substantial opportunities for developing energy management systems that systematically integrate intelligent \emph{prosumers} (both producers and consumers) into the grid, including industrial plants and large buildings.} In smart logistics and intelligent warehousing, the seamless integration of connected goods with mobile robotics and other automation technologies is catalyzing a transformative shift~\cite{liu+23}, with the potential to extend across broader segments of the supply chain. This includes the cyber-physical control of the entire freight transport system, integrating automated trucks and ships~\cite{bess+16}.

The design requirements for autonomous ICPS must effectively navigate several challenging trade-offs among control, communication, and computational capabilities. The core issue in networked control system design is on how to ensure a certain level of control performance despite limitations or uncertainties within the communication and computational platforms. Key research questions include how to design controllers that can manage network imperfections (such as delays, losses, and outages), how to tune communication protocols to meet the demands of real-time control loops, and how to jointly optimize control, communication performance, and resource utilization. Frameworks to address these challenges, particularly in wireless sensor/actuator networks, have been systematically developed and rigorously tested in practice~\cite{park+18}. \R{More recent approaches involves goal-oriented integration of sensing, communication, computing and control~\cite{cao+24}, and regret-optimal cross-layer control designs.}

\R{Many automation and control systems operate under stringent timing requirements. The influence of time delays on closed-loop system performance is a well-established topic in control theory~\cite{frid14}. When considering ICPS, the underlying time architecture and its assumptions become even more crucial. The presence of time delays, jitter, and synchronization can significantly impact system performance~\cite{li+24}. In fact, control systems may lose stability if an adequate amount of data is not consistently circulated. It is possible to precisely determine the required data rate to maintain stability or achieve specific control performance~\cite{nair+07}. Given that sensor or control data can be intermittently lost, estimation and control algorithms have been developed to address issues arising from lossy communication networks~\cite{schen+07}.}

\R{From a communication perspective, these foundational properties of networked control have given rise to the concept of semantic/goal-oriented communications, where the significance of specific data is considered in the network protocol~\cite{uysal+22}. The value of information contained in a data packet heavily depends on its location within the network and the timing of its delivery. Network resources should not be wasted on transmitting outdated data. The urgency of executing feedback control actions has been leveraged in event-triggered control, where sensing or actuation data are transmitted only when necessary. This approach has been extended to various scenarios, such as event-triggered model predictive control. Finally, cyber-physical security and privacy requirements are naturally imposed on networked control systems, as the confidentiality, integrity, and availability of information across the feedback loop are paramount for many real-time and safety-critical control systems.}

\subsection{New-Generation Digital Infrastructure} \label{subsection:Digital_Infrastructure}

An evolving computing and communications infrastructure is currently enabling a digital transformation in ICPS, creating a new-generation digital infrastructure. The foundation of new generation industrial systems builds on distributed computing and newer wired/wireless networking capabilities to support time-critical and other traffic types in a converged network. As computing resources at the edge/cloud take a key role in tasks that require high processing power, such as AI and other computational tasks that involve large amounts of data, the network (both wired and wireless) needs to guarantee lower latency with higher reliability. Avoiding congestion-induced delays and losses is paramount for such converged networks.


TSN technologies have emerged as a toolbox to enable converged networks where mixed criticality applications co-exist. TSN standards defined by the IEEE 802.1 Task Group include features for time synchronization, low latency, reliability, and resource management that can be enabled on top of Ethernet-based local area networks (LAN)~\cite{nasrallah2018ultra}. These features can be combined to build a network infrastructure in which time-sensitive traffic receives priority treatment to guarantee latency bounds with high reliability. Recently, TSN features have also been extended to Wi-Fi~\cite{cavalcanti2022wifi} and enabled over 5G networks~\cite{rost2022performance}. Recent advances in IEEE 802.11 standards, such as new scheduling capabilities in 802.11ax and multi-link operation (MLO) in 802.11be have enabled newer generation Wi-Fi devices and networks that can provide much more granular QoS with lower latency. Combined with TSN enhancements, Wi-Fi~6 solutions have been used to enable end-based robotics and other industrial automation use cases~\cite{sudhakaran2022wirelesstsn}. The latest Wi-Fi~7 chipsets based on the IEEE 802.11be amendment include the capability to operate in multiple links that can be used to not only reduce the channel access latency, but also to increase reliability through redundant transmissions and guarantee low latency performance during roaming~\cite{adame2021time}.

5G standards have also introduced uRLLC capabilities that can be used to support high QoS guarantees for time-sensitive traffic streams. The 5G Release 16 specification has also defined mechanisms to integrate a 5G system with TSN-based networks and enable time synchronization and low latency guarantees in an extended wired-wireless TSN infrastructure. Recent research works on interfacing 5G capabilities with TSN have been demonstrated in~\cite{rost2022performance}.

\begin{figure}[t]
	\centering \includegraphics[trim={0.3cm 2cm 0cm 0cm},clip,width=\linewidth]{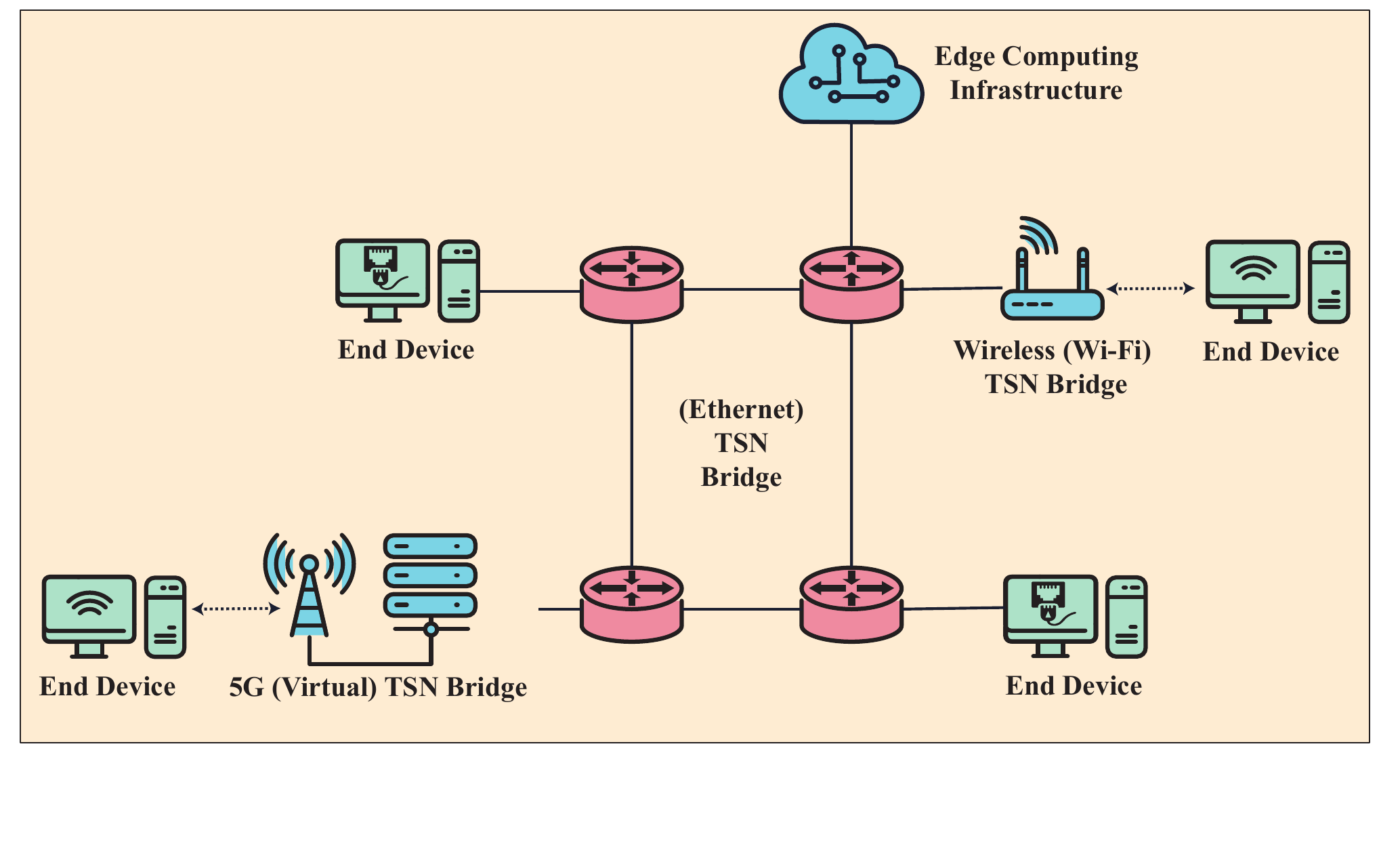}
	\caption{Converged network architecture in TSN-based new-generation digital infrastructure.}
	\label{fig:ConvergedNetwork-Architecture}
\end{figure}

Fig.~\ref{fig:ConvergedNetwork-Architecture} illustrates the envisioned computing and network infrastructure of the future that incorporate TSN and other latency and reliability enhancements in wireless protocols to support distributed computing. Computing resources for time-critical tasks may be available at distributed locations forming an edge computing infrastructure, which is served by a heterogeneous network including Ethernet, Wi-Fi and 5G technologies supporting TSN capabilities. The TSN features can ensure accurate time distribution and guarantee latency bounds for time-critical flows across the network.

\subsection{Closing the Gap with 3C Co-design}

Although there has been significant progress in innovating communication, computing and control technologies to support new ICPS applications and use cases, there is still a significant gap between the current/emerging digital infrastructure and the vision of an autonomous ICPS. The new-generation wired/wireless TSN-enabled digital infrastructure currently underway can only provide almost-deterministic communication at best, and does not effectively consider the adjacent computing and control requirements for achieving autonomous ICPS. Similarly, advancements in computing and control domains do not closely consider and tightly integrate with innovations in the underlying communication infrastructure.

\R{To close this gap, we propose the adoption of Cloud-Fog Automation as the new paradigm to fully realize the vision of autonomous ICPS. In particular, we call for academic and industrial research communities to devote efforts into 3C co-design -- co-designing technologies in communication, computing and control by considering the inter-related ICPS requirements. The central diagram in Fig. 1 visually captures the essence of 3C co-design. 3C co-design is a multi-dimensional synergy concept that is crucial in bridging the divide between current/emerging new-generation digital infrastructure and the realization of a fully autonomous ICPS.} For instance, with the deployment of new wireless technologies to empower almost-deterministic uRLLC, a joint design of optimal control and computing techniques is non-negotiable for time-critical ICPS. It is also imperative that functional safety and security of such systems are not compromised. A prime focus will be on system-wide/application-level performance, driven by goal-oriented communication of interconnecting communication and networking technologies.

\R{In the following sections, we further expand on the aforementioned research challenges and future directions by providing a detailed exploration of the three domains of 3C co-design: (i) goal-oriented communication, (ii) virtualization-empowered computing, and (iii) QoS-aware control.} We subsequently address cyber-physical security as a significant concern that intertwines with each of the 3C domains. Each section offers an overview of the respective area, reviews current and emerging technologies in the literature, and examines their inter-connections within the 3C domains. Each section outlines future directions respective to its own domain.

\section{Goal-oriented Communication }
\label{section3_communication}


Historically, the development of mobile wireless communication systems has been driven by the increasing demand for the transmission of higher amounts of data by individual users, as well as the rapidly growing number of users. Therefore, the design and optimization of communication systems has been targeted at maximizing communication efficiency in terms of coverage costs and the delivery of information bits from a transmitter to a receiver. This is expressed in well-known communication metrics such as spectrum usage efficiency, network capacity, bit error rates, and packet error rates. Significant advances have been made in PHY and MAC layer techniques, such as channel encoding, hybrid automatic repeat request (HARQ) mechanisms, multiple-input multiple-output (MIMO) and beam-forming techniques, improved channel state information (CSI) estimation and signaling protocols, as well as resource allocation and scheduling algorithms. As a result, within this historical optimization paradigm, most of the additional network capacity is now primarily available in very high frequency bands, such as in millimeter bands (mmWave) and beyond.

\R{In the development of the fifth generation (5G) mobile wireless technology, the focus of communication have expanded beyond human-oriented consumption to include machine-oriented communication. With the wireless communication industry's increased interests in IIoT and ICPS use cases, significant research and development efforts have been devoted to minimizing end-to-end latency, which is evident by 5G's uRLLC specification. However, it remains a challenge for wireless networks to address various applications needs and requirements, including bandwidth efficiency, low-latency, and low power/energy efficiency. Furthermore, communication technologies need to be adaptive to networks with very different communication demands, traffic and application patterns, which is often the case in private enterprise networks, and machine-oriented communication in ICPS.}

In recent discussions about the future challenges in the development of wireless communication technology, in addition to further optimizations on PHY and MAC layers (for example with novel AI/ML approaches), there is a growing interest for a more holistic and application-driven optimization at the semantic level, which goes well beyond data bit or packet delivery. The pertinent problem of communicating the meaning of the data was first introduced by Shannon and Weaver in 1949~\cite{shannon1949theory}, where their work divides the broad subject of communication into three levels:
\begin{itemize}
    \item Level A: How accurately can the symbols of communication be transmitted? (\textit{The technical problem.})
    \item Level B: How precisely do the transmitted symbols convey the desired meaning? (\textit{The semantic problem.})
    \item Level C: How effectively does the received meaning affect conduct in the desired way? (\textit{The effectiveness problem.})
\end{itemize}

Most of the works in communication and information theory, design and optimization since the introduction of Shannon's work have been focusing on addressing the \textit{technical problem} on how to deliver bits and symbols of communication. With the introduction of the \textit{semantic problem}, the primary goal of data communication is to accurately deliver the meaning of the data from a sender to a receiver. In many applications, this implies significant reduction of the data amount to be transmitted as semantically irrelevant data can be withheld from transmission. Specifically, when discussing machine-to-machine (M2M) control communication, the meaning of the information is extracted and processed by a control application. As a result, there are natural interests in designing communication mechanisms to progress to Level C and optimize the \textit{effectiveness problem} of communication with the consideration of control goals of the application.

\begin{figure*}[htp]
	\centering
 \includegraphics[trim={0cm 0cm 0cm 0cm},clip,width=0.7\linewidth]{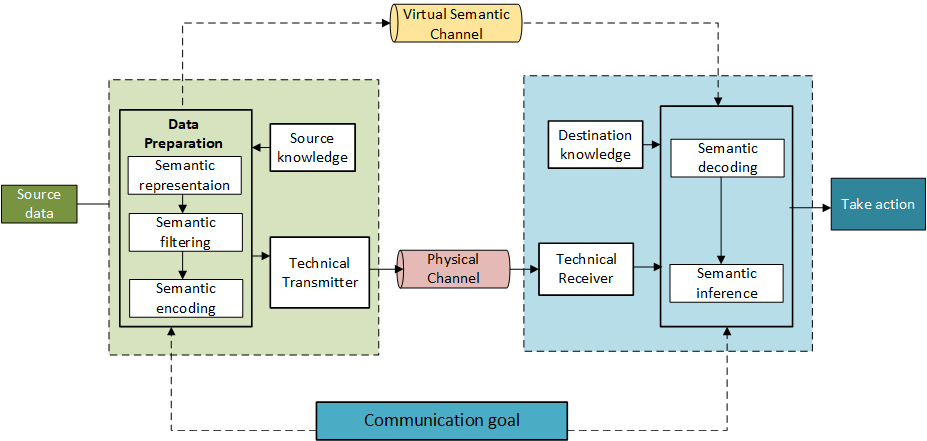}
	\caption{System model for goal-oriented communication (adapted from~\cite{niu2022semshift} and~\cite{getu2023semgoalsurvey}).}
	\label{fig:goal_oriented}
\end{figure*}

Recently, the subset of semantic communication which explicitly considers application goals or tasks is named as ``goal-oriented communication". Fig.~\ref{fig:goal_oriented} presents the general system model for goal-oriented communication. Although it naturally extends semantic communication for machine-type control applications, goal-oriented communication is not limited to such use cases. It can be extensively applied to human-oriented communication. Some research works have started to go beyond communication and include aspects of compute, cache, and sensing jointly in alignment with control application goals~\cite{cao2024gointegration}. We refer the readers to several excellent visionary, overview and survey articles on goal-oriented communication, which appeared in recent years~\cite{calvanese20216Gsemantic,luo2022semmcomm,niu2022semshift,qin2022semanticcomm,gunduz2023beyondbits,getu2023semgoalsurvey}. In the following sub-sections, we specifically highlight several research works and directions with the focus of goal-oriented communication relevant to the ICPS context.


\subsection{Metrics of Goal-oriented Communication}

The first questions raised in the new paradigm of semantic and goal-oriented communication are on defining suitable metrics, such as how to formulate and measure the meaning of information, how to determine the communication goals, or how to define communication effectiveness. These key questions have been raised by the research community since the conceptualization of goal-oriented communication~\cite{getu2023semgoalsurvey}.

\R{\textit{Age-of-Information (AoI)} is one of the most important and novel metrics~\cite{kaul2012statusupdates,yates2021aoi}. As its name suggests, AoI indicates the time passed since the latest and not yet received data has been generated at the source (e.g., from a sensor). Although it does not have explicit semantic values, it indicates the freshness of information through its lifetime, which especially for ICPS applications often implies that fresher information is much more relevant and critical to act upon~\cite{abd-elmagid2019aoiiot}. Higher AoI values translate to higher delays, where precise control and decision-making may not be possible. Time-critical ICPS applications can benefit from the integration of AoI at the semantic level to assist with task prioritization and resource scheduling. AoI has been already been used in several research works, focusing on the co-design of control and communication systems in ICPS~\cite{wang2021aoicontrol,mamduhi2020aoicontrol}.}


Since the first introduction of AoI, there has been many adaptations to the AoI metric~\cite{getu2023semgoalsurvey,salimnejad2024ageinformationversionssemantic}. In~\cite{kosta2017voi}, the authors proposed \textit{Value-of-Information} which extends AoI with a non-linear utility function to reflect the sensitivity of applications to the freshness of the information. With a similar motivation, the work in~\cite{zheng2020uoi} introduced \textit{Urgency of Information}, a novel metric which multiplies time-dependent context-aware weights and a cost function for the use of delayed information. The system urgency status for a new information update is characterized and used for data sampling at the source. It is also used for scheduling when multiple users share the same network resource. This intuitively makes sense, for example, when the system gets closer to certain critical threshold it is not allowed to cross, the urgency for the update increases. On the other hand, if the system is far away from a critical threshold, it is more delay-tolerant. The authors specifically investigated the performance benefits in a remotely control cart-pole systems (a mobile inverted pendulum setup), which accurately captures the control challenges in industrial control applications. In a similar manner, the authors of~\cite{agheli2024goe} introduced \textit{Grade of Effectiveness} metric, which incorporates both freshness and usefulness attributes of the information.

An important innovation in the AoI metric for goal-oriented communication is the direction of information flow. AoI, as originally defined, is uni-directional by its nature and can be used either to optimize the downlink (DL) or uplink (UL) direction separately. In contrast, in control systems, the cycle from sensor to controller and then to the actuator needs to be taken into account as an end-to-end metric. The authors of~\cite{desantana2021aol} proposed the adapted \textit{Age-of-Loop} metric to address this particular issue with AoI, and demonstrated the benefits in terms of control accuracy, where both DL and UL links are both jointly considered. In~\cite{li2024semcommetrics}, the authors provided a good summary and comprehensive comparison of various AoI-adapted metrics in multi-directional information flow scenarios. They also introduced a more complex multi-dimensional tensor metric to unify multiple relevant aspects~\cite{li2024gotesnsor}.

\subsection{Goal-oriented Communication Design}

In the new goal-oriented communication paradigm (as presented in Fig.~\ref{fig:goal_oriented}), nearly all aspects of communication can be re-visited and aligned with specific communication tasks with new optimization metrics.

\subsubsection{Data preparation}
One of the initial important questions of goal-oriented communication is deciding which specific data needs to be send. Multiple goal-oriented and semantic optimizations can be undertaken at the source of the data itself, such as data quantization~\cite{zou2023goquant}, data sampling~\cite{pappas2021gotracking}, data compression~\cite{zhang2022goiot}, and source coding~\cite{agheli2024semfiltering,ginenez-guzman2024gosourcecoding}. With such measures, the goal-oriented communication system can better reduce the amount of data sent when the data is less critical for the communication goal; or increase if the controlled system becomes sensitive to sensor information and updates.

\subsubsection{Joint source-channel coding}
Next, the question on how to transmit data in a goal-oriented and semantic manner needs to be addressed. With advances in deep learning, researchers have started to investigate joint source-channel coding -- two processing tasks which are usually performed separately from each other due to their complexity. With the introduction of semantic communication, a joint integration of channel coding (technical transmitter in Fig.~\ref{fig:goal_oriented}) and semantic encoding can be undertaken. \R{Further, recent works have demonstrated that joint source-channel coding can be approached with deep learning techniques and it brings significant benefits for specialized applications (e.g., video transmission, image sensor data), which use joint data coding/protection that are aligned with wireless channel dynamics~\cite{xu2023deepjscc,lo2023semedge,tung2022deepwive}.}


\subsubsection{Resource allocation}
Due to the bandwidth constraints in wireless networks, the allocation of wireless resource becomes critical for reaching application goals. With recent advances in reinforcement learning (RL) algorithms, several research teams have devoted their efforts to apply RL methods to address the resource allocation problem with goal-oriented  communication~\cite{tung2022deepwive,ceran2023rl_aoi,mason2024marcomml,talli2024pragmaticcommunicationremotecontrol}. These works include the use of multi-agent RL for multiple network users such as sensors and robots. It is demonstrated that such learning algorithms are able to provide specialized action policies in complex dynamic network scenarios, where classical scheduling methods approach their limits. In~\cite{pezone2022goalorientedcomm}, the authors extended the analysis to learning the effective allocation of communication and computing resources.

\subsubsection{Communication protocols and emergent communication}

Finally, the protocol aspects for coordinating data transmissions from multiple agents can be optimized according to the design of goal-oriented communication systems. In~\cite{tung2021effcomm, talli2024pushpullbasedeffectivecommunication}, the authors have adopted an approach similar to multi-agent RL to investigate the optimization of pull and push-type protocols for certain control applications. In a more generalized approach, emergent communication among agents (which take actions), interact with environment and use communication to coordinate and cooperate so that their goal-oriented reward is maximized. This approach is investigated with RL frameworks in~\cite{foerster2016learning2communicate,pmlr-v97-das19a,chafii2023emergent}. These works represent early steps in this direction, and demonstrated the possibility and feasibility for the learning-based emergence of coordination and communication protocols for dedicated tasks such as industrial task offloading in ICPS~\cite{mostafa2023emergentcomm}.

\subsection{Future Directions}


While the rapidly evolving field of goal-oriented communication aims to overcome the limitations of current communication technologies, it remains in its early stages and faces many challenges and unresolved questions. Here we outline several key future research directions.

\begin{enumerate}
\item {\it Unified Theory}. Compared to conventional information theory, it is still challenging to derive a unified theory for semantic and goal-oriented communication to fully understand the fundamental limits and constraints of this new paradigm.

\item {\it Metrics}. There are many metrics discussed in the literature, which actively contributed to the the design goal-oriented communication. However, these metrics are difficult to generalize, and they only address some aspects of goal-oriented communication aspects and limited to specific scenarios.

\item {\it Practical implementations}. To date, most of the research works in the field of goal-oriented communication are based on simulations or analytical frameworks. However, it is important to investigate the practical implementations of goal-oriented communication, and experimentally demonstrate the realized benefits in real hardware and wireless channels.

\item {\it Legacy co-existence}. Even when proved beneficial in some scenarios, new goal-oriented communication systems are likely to co-exist with conventional networks and share their resources~\cite{merluzzi2024golegacy6g}. The co-existence with legacy systems will bring about additional challenges in terms of performance benefits and trade-offs. Standardization efforts are also crucial in ensuring that these system designs are safe and practical for mass deployment.

\item {\it Beyond communication}. The paradigm of goal-oriented communication have started to expand beyond communication. New metrics can be used to take into consideration other applications with relevant resources such as compute, caching, or wireless sensing. This trend has also motivated the need for an intrinsic 3C co-design in the context of ICPS, with a particular focus on industrial control applications.
\end{enumerate}

Finally, it is imperative to further extend and clearly demonstrate the range of applications that could potentially benefit from goal-oriented communication. Currently, there is a particular focus on video, AR/VR and control-based applications. Further expansion, investigation and specification of various new and emerging applications are required from both academia and industry research.

\section{Virtualization-Empowered Computing}
\label{section4_computing}




\R{Virtualization represents a fundamental computing principle, where the underlying hardware resources are logically partitioned/shared, and presented to the upper-layer software as if it had exclusive hardware access. It enables application scalability and the consolidation of multiple workloads over the same physical hardware~\cite{Functionalities2015, pearce2013virtualization}. Since the off-the-shelf availability of hypervisors and virtual machines (VMs) two decades ago, virtualization has been rapidly adopted across various sectors. In particular, the cloud computing industry has widely adopted virtualization as its key enabling technology to drive the sharing/renting of computing resources to multiple clients. As the technology mature, virtualization has progressively proliferated ICPS and embedded system sectors, which possess strict functional safety and dependability requirements~\cite{li2017industrial}.}


\R{Innovations in ICPS and computing architectures such as fog/edge computing have demonstrated the feasibility of leveraging virtualization technologies in commercial off-the-shelf multi-core processors in industrial environments with the goals of achieving system-wide scalability, reliability, availability, security, durability, portability, and integration with legacy systems~\cite{Pop2021,Mahmud2014}. However, despite much efforts in academic and industrial research, there are still significant challenges in adapting information technology (IT)-specific virtualization mechanisms towards hardware/software stacks in operational technology (OT)-based ICPS applications.}

\begin{figure*}[htp]
	\centering \includegraphics[trim={1.5cm 2cm 2cm 0cm},clip,width=0.9\linewidth]{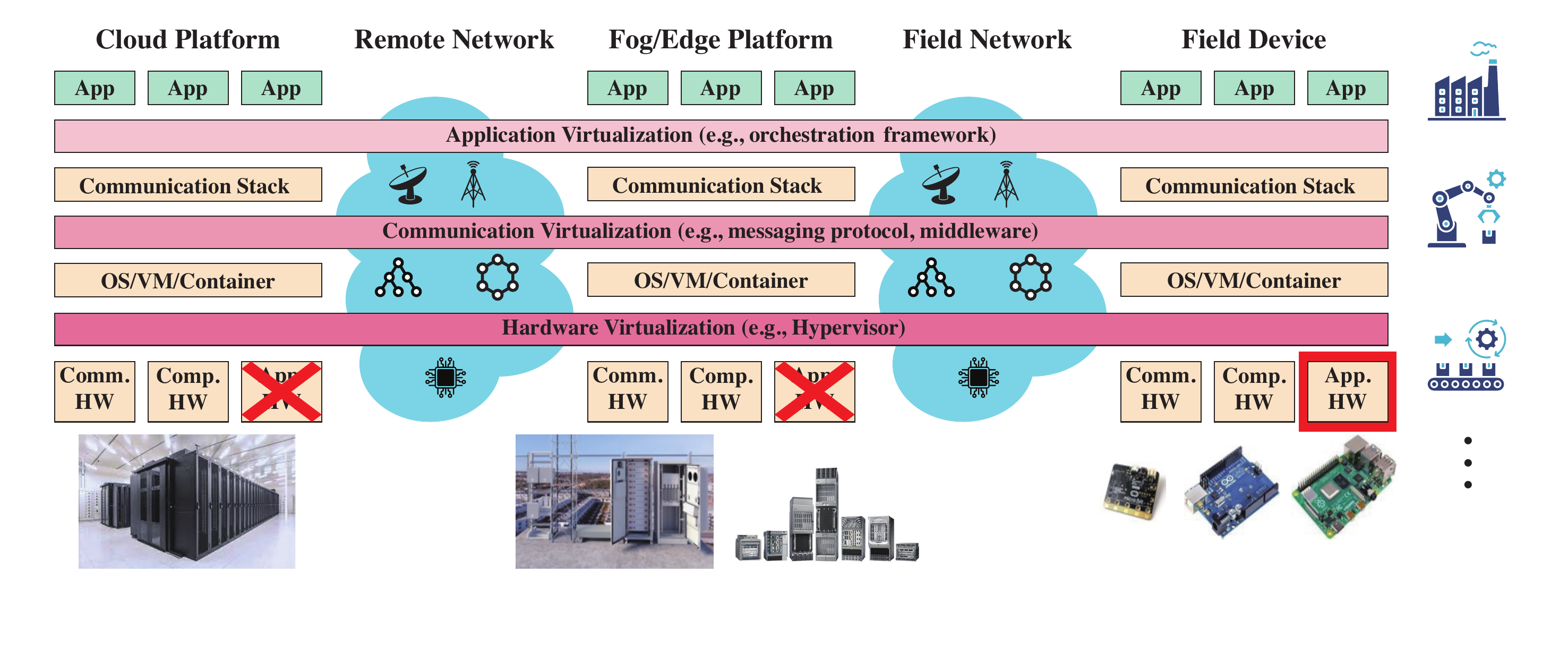}
	\caption{\R{Virtualized-empowered computing across application, communication, and hardware layers for executing control logics within autonomous ICPS.}} 
	\label{fig:three_pillars}
\end{figure*}

Fig.~\ref{fig:three_pillars} illustrates the role of virtualization technologies in ICPS, which can be delineated into three distinct yet inter-connected layers: (i) \textit{application}; (ii) \textit{communication}; and (iii) \textit{hardware} virtualization. These virtualization layers span from the field devices, field network, fog/edge platforms, remote network, to the cloud platform. These inserted virtualization layers enable the ``lower-layer independency", which protects the business interests of the vendors, however, they also incur some ``black holes" of overheads.


\R{The notion of \textit{`mixed-criticality'} in virtualization is paramount in ICPS~\cite{Cinque2022}. In cases where tasks with different safety-criticality are executed on separate VMs, a high level of temporal and spatial isolation between VMs are required to guarantee their safety and security.} In cases that require dynamic reconfiguration of networks and migration of tasks across virtualized computing nodes, the safety of VM migration needs to be ensured so that the timing and execution behaviors are not adversely affected~\cite{Gundall2020}.

The aforementioned stringent industrial requirements led to the development of the concept of \textit{`deterministic virtualization'}~\cite{Ruh2019}, which specifies: (i) \textit{determinism} in guaranteeing predicable latencies and low jitter; (ii) \textit{safety certification} of VMs that run safety-critical control tasks; (iii) \textit{isolation} of VMs to ensure system security and safety, and reduce non-deterministic timing originating from shared resources; and (iv) \textit{flexibility} in replacing VMs and task sets during runtime for scalability, re-configurability and safe live migration.

\R{While virtualization technologies have significantly matured and widely adopted at the hardware layer, it has recently found its way into the application and communication layers, particularly in ICPS. Hence, their tight interlocking and interactions with hardware virtualization for achieving deterministic virtualization merits the community's attention. Below we present a concise review of the state-of-the-art in ICPS-specific application, communication, and hardware virtualization.}


\subsection{Application Virtualization}

\subsubsection{OS-level virtualization}
In contrast to the high performance overhead incurred by hardware virtualization (where the guest OS is executed inside a VM), OS-level virtualization using containers are much lighter weight and could promisingly address the stringent footprint requirements of real-time embedded systems in ICPS~\cite{Moga2016}. The key innovation of containers is the abstraction of OS processes, isolating the user-space while sharing the host OS kernel (e.g., Linux uses the \textit{namespace} process isolation mechanism, and \textit{cgroups} to provide resource management capabilities). Containers are virtual environments with local APIs, libraries and settings, with their own perception of virtual resources (e.g, CPU, memory, file systems, network interfaces, etc). The container start-up/shut-down times are minimal, and it can be built from templates and shared/extended/reused across different containers. Entire applications could be containerized and spread across different containers, making containers a versatile and light-weight solution for dynamic environments such as task scheduling and resource allocation in ICPS~\cite{Member2018}.

\R{The adoption of OS-level virtualization in ICPS is a recent trend. The aim is to leverage containers instead of VMs to achieve isolation with minimal overhead while meeting the performance requirements of mixed-criticality systems~\cite{Barletta2024}. Experimental studies and analysis have demonstrated the feasibility and benefits of using container and its in-built orchestration frameworks (e.g., Docker Swarm) and platforms (e.g., Kubernetes) in ICPS~\cite{Moga2016}. For example, the work in~\cite{goldschmidt2018container} presented an architecture for a container-based multi-purpose industrial controller, including support for real-time PLC execution engines and emulation of legacy engines. The work in~\cite{Sollfrank2021} evaluated the time delay impacts of the Docker software with a real-world use case, and demonstrated that Docker containers can be used as a distributed lightweight virtualization scheme to meet the soft real-time requirements of industrial automation. In~\cite{morabito2017evaluating}, the authors evaluated the performance of container-based solutions for service provisioning in wired/wireless industrial clusters for improved manageability and scalability. A recent work in~\cite{Becattini2024} proposed a containerized application-driven service architecture for digital twin networks using Kubernetes in industrial networked environments, and provided empirical application insights for a practical extension of the ITU-T Y.3090 standard.}


\subsubsection{Microservices}

\R{Microservices is a software development approach that structures an application as a collection of loosely coupled and independently deployable services~\cite{thones2015microservices}. Each microservice fulfills a specific business requirement, encapsulating its own logic, data, and communication interfaces. This architecture enhances system modularity, scalability, and fault isolation. Consequently, it enables teams to develop, deploy, and scale different parts of the system independently. Microservices typically communicate via lightweight protocols such as HTTP or messaging queues, and often deployed in tandem with containerization technologies. Microservices are suitable for complex, distributed systems where agility, continuous delivery, and resilience are critical. Building upon its success in the computing industry, research and development into ICPS-specific microservices and their orchestration frameworks are emerging, creating new opportunities for innovations and advancements~\cite{dragoni2017microservices}.}


\R{IEC 61499 is a notable standard for system-level microservice orchestration. The work in~\cite{Dai2023} proposed a microservices-enabled orchestration architecture for deploying features specified in IEC 61499, with containers to provide maximum flexibility, interoperability, and efficiency for industrial edge applications. It maps the rules between IEC 61499 function blocks and microservice architecture for both the design time and runtime. In~\cite{thramboulidis2018cyber}, the authors proposed a cyber-physical microservices framework for manufacturing based on unified modeling language (UML) models. Guidelines for functional requirement decomposition are formulated in~\cite{homay2020service} and modeled across the process, machine, and functionality layers of the process control systems. A Kubernetes microservices approach is proposed in~\cite{koziolek2021dynamic} to encapsulate containers, with a special web transfer service for exchanging data between virtual PLCs (vPLCs). A cloud-based IEC 61499 virtual commissioning method is presented in~\cite{lyu2021towards} to orchestrate a set of application runtimes with microservices.}


\subsection{Communication Virtualization}

\subsubsection{Software-defined Networking (SDN) and Network Function Virtualization (NFV)}

Communication virtualization is primarily driven by the virtualization of the network layer, with technologies such as SDN and NFV. SDN separates the control plane from the data plane, which were deemed inseparable in proprietary hardware/software in the past. The SDN control plane makes traffic-handling decisions, while the data plane solely moves data content between two network nodes. SDN enables a global view of the network, and achieves hardware independence while providing intelligent policing/routing decisions, and ensuring redundancy and fault tolerance. NFV~\cite{Han2015} is a resulting architecture that specifies the executing of SDN functions independent of any specific hardware platform. It is an infrastructure that administers and orchestrates virtual network functions (VNF), which are individual network services running as a software-only VM instances on generic hardware~\cite{Li2024}.

\R{SDN/NFV have been deployed extensively in data center networks but have only recently realized in closed and controlled ICPS networks~\cite{Leonardi}. Recent research efforts in designing/deploying centralized~\cite{farris2018survey} and distributed SDN controllers~\cite{bannour2017distributed} for network automation (including sensing/actuating, process orchestration/automation, etc) have demonstrated the applicability of SDN/NFV in ICPS~\cite{Industry2023}. However, the timely synchronization of resource demand information of VNFs with the SDN controller is critical as the delays directly impact the QoS of emerging applications (e.g., industrial augmented reality, digital twins, human-robot collaboration) relying on such information. Hence, the authors in~\cite{Tang2024} proposed a SDN/NFV network slicing architecture, with ML-based algorithms for efficient VNF mapping and scheduling. To address the challenges of VNF placement, the work in~\cite{Liang2023} formulated VNF placements as one-shot/NP-hard problems and proposed a heuristic online algorithm, whereas optimization modulo theories (MaxSMT) with expressive constraints and formal verification are proposed in~\cite{Marchetto2021}.}


Several recent works have focused on building experimental platforms that implements joint radio resource allocation and beam-forming with SDN/NFV~\cite{Rahimi2022}, supports tactile Internet with novel 5G/6G uRLLC and blockchain-based methodologies in smart manufacturing~\cite{Mekikis2020, Huang2021}, and enables a resilient network with a pseudo-honeypot strategy~\cite{Du2020}. Network slicing management with remote adaption and orchestration~\cite{Ji2022}, transfer learning~\cite{Mai2022} and graph neural networks~\cite{Wang2022a} are also recent research trends.

\subsubsection{Virtual Network Embedding (VNE)}

Network virtualization technologies such as NFV rely on algorithms that can instantiate virtualized networks on the infrastructure to optimize the layout of service-specific metrics/indicators~\cite{fischer2013virtual}. Algorithms that dictate these behavioral properties are referred to as VNE. VNE algorithms solve virtual network requests and efficiently allocate physical node/link resources. Since VNE attempts to optimize for multiple constraints, including attributes such as CPU, storage, bandwidth, request admissions, and other dynamic requirements, it is classified as an NP-hard problem and could benefit from AI/ML-based strategies~\cite{xiao2023dvne}.

\R{VNE algorithms governs the elasticity of virtualized node and link resources in ICPS networks, and it is critical that VNE satisfy the multi-dimensional QoS requirements. For instance, the work in~\cite{Cao2020} proposed a dynamic embedding scheme to optimize performance of each service node, and as each node satisfies the QoS requirements, a re-embedding scheme of the heuristic algorithm will be performed. This scheme ensures the flexibility of virtual nodes and links assignment while fulfilling the customized QoS requirements. In~\cite{li2019intelligent}, the authors proposed an intelligent latency-aware VNE scheme to provide deadline guarantees for industrial virtual networks by implementing both static embedding (with an any-path algorithm) and dynamic forwarding (with deep-learning). Application-driven VNE approaches with an any-path routing scheme are proposed in~\cite{li2017application}. These approaches exploit unique features in wireless channels in factory automation systems.}


\subsection{Hardware Virtualization}

\subsubsection{Virtualized PLC (vPLC)}

vPLC is introduced to address the limitations and challenges posed by physical PLCs~\cite{km-iotops-iiot-frwk-02}. vPLC is a hardware-agnostic abstraction of the control unit and memory functions of a PLC, however, it still requires an interface to communicate with the I/O modules. vPLC is able to incorporate multiple control functions and provide benefits in terms of processing capabilities, flexibility and efficient resource use interoperability and optimization between different I/O module vendors, reducing device density in factory floor, and cost savings. It integrates both IT and OT software components, and plays a critical role towards an integrated and autonomous ICPS~\cite{azarmipour2019plc}.

\R{A recent work in~\cite{Gaffurini2024} presented a methodology for comprehensively evaluating the communication performance of vPLCs (based on a set of defined metrics from corresponding transaction round trip time, and the transmission latencies between devices) when exchanging data for supervision, coordination, and control with other machines and supervisory control and data acquisition (SCADA) in machine-to-machine (M2M) scenarios. A comparison platform is designed to compare against real PLCs, along with the creation and validation of an analytical model with full network traffic access, which can be used for simulators for worst-case analyses. Experiments indicated that vPLC could work as fast as a real PLC with minimum communication latency, but could potentially induced randomized delays (similar to jitter) due to the Internet Protocol (IP) stack implementation of the vPLC.}


\subsubsection{Real-time hypervisors}
\R{Real-time hypervisors add explicit support for managing the time budged allocated to VMs with specific scheduling algorithms, to assure that individual VMs comply with the stringent and explicit timing constraints~\cite{Scordino2020}. Dynamic hypervisors map VM-related resources as needed, whereas static hypervisors provides a one-to-one mapping between virtual and physical hardware resources. Static solutions have lower performance overhead and they are more robust to failures and mis-configurations, hence they are often embedded in safety-critical and mixed-criticality applications~\cite{Cinque2022}. The work in~\cite{Networks2020} proposed a dynamic VM migration approach with a novel online scheduling algorithm based on minimal distance tree - heuristic breadth first search (MDTC-HB) construction, in multi-cast TSN networks to guarantee the firm latency of control messages. Experiments demonstrated the efficacy of the framework in terms of rapid response time while only consuming 50\% bandwidth. In~\cite{Orciari2024}, the authors proposed a virtualized platform that specifically address the hard real-time requirements of advanced control mechatronic systems. The platform uses the Jailhouse hypervisor and designates micro-controllers with specific/routine computational tasks, and offloads heavy/hard real-time computations to micro-processor cores, which is analogous to the recommendations provided in~\cite{Cilardo2022}.}


\subsection{Future Directions}
There are many exciting future research directions in virtualization-empowered computing to support 3C co-design towards autonomous ICPS. We present several promising directions below.

\subsubsection{Application virtualization to support TSN}
Although advances have been made in designing deterministic execution of control and computing tasks in virtualized environments, typical use cases are still self-contained in static contexts, and originates from cloud computing~\cite{Ruh2019}. Therefore, future work will need to rethink a new paradigm for real-time hypervisors, containers and their orchestration frameworks to support deterministic virtualization with a particular focus on highly granular time-sensitive ICPS networks such as TSN, which includes implementing a deterministic I/O, dynamic, coordinated VM, task and network schedules for flexible yet safe control in virtualized platforms.

\subsubsection{Knowledge-centric networking (KCN) with SDN/NFV}
KCN is a new and exciting field of research that emerged from content/information-centric networking (CCN/ICN), which leverages in-network computing, in-network storage, and in-network communication to create the needed knowledge, which allows sensing devices to obtain useful data from anywhere in the network~\cite{Wu2019}. Much of the related work are performed in the public Internet networks, and research efforts on KCN in ICPS networks are limited. With many communication virtualization technologies already driven by SDN/NFV, KCN is a new kid-on-the-block that could significantly improve network analytics and intelligence in ICPS, providing a complete framework from front-end sensing to back-end networking.

\subsubsection{AI/ML for virtualized system testing and certifications}
Industry standard compliance is paramount when deploying virtualization in ICPS. The wide variety of hardware virtualization approaches and their complexity present a challenge in ensuring standard compliance~\cite{Cinque2022}. For example, emerging approaches such as micro-kernels provide strong security and effective verification, but still yet to be fully safety-guaranteed when deploying in industrial environments. There is also a lack of benchmarks and effective test suites to produce evidence in the safety certification process. Therefore, future research could focus on leveraging AI/ML to ensure the compliance of safety and security guidelines through rigorous robustness and performance testing processes, while also informing future development and evolution of industry standards.

\section{QoS-aware Control}
\label{section5_control}



QoS-aware control is a critical component in the design and operation of ICPS, where precision, reliability, and timeliness are fundamental to the success of autonomous industrial operations. \R{In ICPS, the convergence of industrial control and computational systems necessitates a tight coupling between communication networks and control mechanisms.} These systems rely on the ability to manage diverse processes in real time, ranging from machinery control to sensor data analysis, hence requiring sophisticated control strategies to ensure system stability and performance under strict QoS requirements. QoS-aware control mechanisms are designed to guarantee these requirements, ensuring that factors such as bandwidth, latency, reliability, and availability are optimized to support time-sensitive and mission-critical industrial processes. Without proper QoS management, the dynamic and complex nature of ICPS can lead to performance bottlenecks, performance degradation, and even system failures that could dramatically disrupt production and compromise safety~\cite{Zhang7498684}.


\R{In the Cloud-Fog Automation paradigm, QoS-aware control is critical for achieving determinism in communication, computation, and control. Since Cloud-Fog Automation is a network-centric architecture where cloud and fog resources collaboratively handle the computational and control requirements of ICPS, it creates a distributed yet synchronized ecosystem. In this architecture, fog nodes located closer to the physical processes in the industrial environment can handle tasks that require low latency and high reliability, while the cloud offers scalable computational power for data-heavy tasks such as predictive maintenance, system optimization, and large-scale data analytics~\cite{BOTTA2016684}. By leveraging this hybrid cloud-fog environment, QoS-aware control ensures that computational resources are efficiently allocated and that control actions are performed within the strict time constraints required by industrial operations. This approach minimizes latency, optimizes bandwidth usage, and ensures real-time decision-making, which are essential for maintaining the operational integrity of autonomous ICPS~\cite{Satyanarayanan7807196}.}

A key challenge in QoS-aware control within ICPS is the need to manage dynamic, unpredictable workloads while ensuring the system meets pre-defined performance thresholds. Traditional control systems, which are often built on static models, struggle to adapt to the dynamic nature of industrial processes where workload fluctuations, network variability, and potential system failures are commonly observed. In contrast, modern QoS-aware control mechanisms employ adaptive strategies that continuously monitor system performance and adjust resources in real time. These systems use feedback control loops, where control decisions are based on real-time data from sensors and actuators across the entire ICPS. Techniques such as model predictive control (MPC) are used to anticipate system behavior and adjust control actions accordingly to prevent QoS violations~\cite{MAYNE2000789}. Furthermore, ML models are increasingly integrated into these control frameworks, enabling predictive QoS management where potential performance degradation is identified before it occurs. This predictive capability is especially valuable in ICPS, where system downtime or delays can lead to costly disruptions in industrial processes~\cite{Lv9617778}.

Security and reliability of QoS-aware control in ICPS are growing concerns as industrial systems become more connected and complex. Cyber threats and system vulnerabilities can have disastrous consequences in ICPS, making security-aware QoS management a critical area of research. Emerging technologies such as blockchain are being explored for their potential to enhance the transparency and integrity of QoS data in industrial environments, providing an additional layer of protection against data tampering and unauthorized access~\cite{BOTTA2016684}. Additionally, advanced communication technologies such as 5G are expected to play a significant role in enhancing QoS by providing ultra-reliable low-latency communication channels that are crucial for real-time control. By combining these technologies with sophisticated QoS-aware control mechanisms, the future vision of autonomous ICPS can be realized with greater reliability, security, and efficiency, and paves the way for new industrial applications~\cite{Satyanarayanan7807196}.

\subsection{State-of-the-art of QoS-Aware Control}

\R{QoS-aware control is vital for achieving the high-performance requirements of autonomous industrial operations. The control mechanisms in these systems must handle complex, real-time interactions between machines, sensors, and actuators, while accounting for communication delays and computational constraints.} Unlike traditional control systems, QoS-aware control takes into account not only the control objectives but also the communication and computational infrastructure that supports it. Within this framework, latency-aware control has emerged as a prominent approach to manage time-sensitive processes. Latency is often the most critical QoS metric in ICPS. Delays in feedback or actuation can cause performance degradation or even system failures. Latency-aware control strategies, such as time-triggered control (TTC) and event-triggered control (ETC). TTC and ETC are particularly effective in managing communication-induced delays by synchronizing control actions with predictable communication intervals. TTC schedules control updates at fixed times, ensuring that control signals are always sent within a pre-defined time frame, whereas ETC dynamically adjusts control actions based on the system's state, hence reducing the communication load when the system is stable and increasing the load during critical changes.

The integration of communication and control is essential for managing the strict timing and computational demands of industrial processes. MPC is one of the most widely used methods in this context, and has been adapted for latency-aware environments~\cite{Liu10507194}. In latency-aware MPC, control actions are calculated not only based on the current state of the system but also with an anticipation of potential communication delays and computational constraints. This anticipatory control framework allows the system to maintain optimal performance even when data transmission or computation takes longer than expected. In addition, latency-aware MPC can be combined with resource management techniques, where computational resources are dynamically allocated to processes that are most sensitive to delays, thus ensuring that critical operations are prioritized~\cite{Zhu10234133}. This approach is often implemented in cloud-fog architectures, where the fog layer handles latency-sensitive tasks closer to the operational environment, while the cloud is responsible for less time-critical, computationally intensive processes. This distribution of computational and control tasks between cloud and fog layers ensures a balanced workload, and reduces the risk of latency-related QoS violations in ICPS.

Another key aspect of QoS-aware control is the co-design of communication and control strategies~\cite{Wang10005612}, which addresses the direct interaction between communication delays and control performance. In traditional control systems, communication and control are often treated independently, leading to sub-optimal performance in networked environments. However, control strategies are designed to explicitly account for the underlying network's behavior in QoS-aware ICPS, including variable communication delays, packet loss, and bandwidth limitations. A typical example is the networked control systems (NCS) framework, where control decisions are made based on the available communication bandwidth and network conditions. In this context, ETC is particularly effective as it reduces unnecessary communication by only sending updates when a significant change in the system state occurs~\cite{zhao9031558}. ETC-based systems is able to dramatically reduce communication overhead while maintaining high control accuracy, making them an ideal candidate for industrial processes that require tight control but operate over bandwidth-limited networks. These techniques ensure that control systems are not overloaded with frequent updates, which could introduce unnecessary delays, but instead are triggered by meaningful events that impact the process.

Furthermore, the application of adaptive and ML-driven control in latency-sensitive environments has become a critical focus in the evolution of QoS-aware control for ICPS~\cite{Wang10024169}. Adaptive control methods are increasingly being employed to dynamically adjust control actions based on changing system states and varying network conditions. For example, latency-aware adaptive control (LAAC) continuously monitors the network and computational latency to adjust control actions in real time, allowing the system to adapt to fluctuating conditions~\cite{lyu10586839}. ML techniques have been integrated into ICPS to enable systems to learn optimal control strategies over time, particularly in environments where the relationship between communication delays and control actions is too complex to model with traditional methods. These learning-based models can optimize control strategies dynamically by observing how different control actions affect system performance under various latency conditions, thus enhancing both responsiveness and efficiency. For instance, RL-based methods can be effectively applied to automatically adjust control update rates in response to network conditions, hence ensuring that communication resources are used efficiently while adhering to strict QoS requirements.

\subsection{QoS-aware Communication and Control Co-design}

Through the lens of 3C co-design, QoS-aware control aims to integrate communication, computing, and control in a unified manner to achieve optimal performance in complex systems~\cite{Zhou9544089}. One key technique is the joint optimization of communication and control systems. This method ensures that control systems are aware of communication delays and packet losses, while communication protocols are designed to prioritize the timely delivery of control signals. Research shows that co-design frameworks such as MPC combined with NCS can help achieve this balance. This joint optimization framework ensures that neither the control objectives nor the communication performance is compromised, which is essential for real-time applications in autonomous ICPS.

Another crucial aspect is the dynamic adaptation of control strategies based on real-time feedback from the communication network~\cite{lyu10586839}. Control algorithms should adjust their operations depending on the current state of the network, such as adjusting control update rates based on available bandwidth or reducing the control task frequency when network delays occur. Techniques such as adaptive control systems or ETC have proven to be effective in adjusting the frequency of control commands based on network conditions. This adaptability ensures that the system maintains stability and performs optimally, even in constrained systems. The ability to react to varying QoS metrics such as latency, jitter, and packet loss in real time is critical for control systems deployed under the Cloud-Fog Automation paradigm, where network conditions are not always predictable due to industrial services being deployed in a distributed fashion.

Efficient resource allocation also plays a vital role in ensuring that QoS-aware control systems operate effectively~\cite{lyu10558844}. In distributed control systems, resources such as bandwidth and computational power are limited. Therefore, resource management techniques that prioritize critical control actions over less important data transmissions are essential. Priority-based resource allocation methods, such as those that use optimization algorithms to assign bandwidth to the most time-sensitive control signals, have been shown to improve system performance~\cite{Martinez9194714}. In networked industrial systems, where communication congestion can significantly impact control accuracy, effective resource management ensures that communication delays are minimized, allowing control loops to function correctly. This dynamic allocation is especially important in large-scale systems, where many devices must share limited communication and computational resources.

Finally, real-time feedback loops and cross-layer designs are crucial for ensuring optimal performance in QoS-aware control systems. Real-time feedback allows control systems to adjust their operation based on current communication performance, hence ensuring responsiveness to changing network conditions. Cross-layer design further enhances the system by integrating control requirements directly into the communication protocols, making the communication system more responsive to control demands. Research in cross-layer optimization techniques, which includes integrating physical and network layers, demonstrates that aligning control and communication objectives leads to more efficient operation in large-scale ICPS. The interplay between control, communication, and computing ensures that all three domains operate in harmony, leading to enhanced performance and stability in future autonomous ICPS.


\subsection{Future Directions}

There are currently several key research directions within the QoS-aware control domain, which will form critical enablers for 3C co-design in autonomous ICPS.

\subsubsection{Adaptive control under communication variability}
Autonomous ICPS will heavily depend on advanced QoS-aware control strategies that ensure seamless interaction between control, communication, and computing. The primary challenge for QoS-aware control lies in ensuring that control systems can function optimally despite the inherent variations in communication and computing capabilities across cloud and fog environments. Control algorithms need to be adaptive and intelligent. They need to respond to changes in communication delays, computational loads, and system dynamics in real time~\cite{Wan9454409}. The challenge is that QoS-aware control must manage the timing and quality of the information exchange between distributed systems while also ensuring that control actions meet the strict performance criteria required for industrial automation.

\subsubsection{Real-time control coordination with communication and computing}
QoS-aware control not only maintains the stability and precision in control loops but also ensures that control decisions are made in coordination with communication latency and computing resources~\cite{Cai9791115}. As the cloud-fog infrastructure handles distributed tasks, the real-time nature of control actions becomes even more critical. Latency and computation delays must be accounted for in control strategies, ensuring that key decisions are made within the timing constraints of the system. For example, real-time control systems operating in environments such as manufacturing or process automation need to adjust dynamically to fluctuations in network performance and computing availability, while simultaneously avoiding system slowdowns or inefficiencies.

\subsubsection{Scaling QoS-aware control for large-scale ICPS}
The interplay between control and communication becomes especially important when ICPS need to scale, which involves a large number of interconnected devices and sensors~\cite{Hu5560641}. QoS-aware control must leverage distributed control architectures to scale, where control decisions are decentralized and localized to reduce communication delays. However, the co-design of control with communication protocols ensures that vital information is still shared across the system in a timely manner, without overwhelming the network. For instance, control algorithms must prioritize critical data flows while ensuring non-essential data does not interfere with time-sensitive control decisions. This coordination allows industrial systems to operate smoothly, thus maintaining stable system performance and avoiding bottlenecks due to communication constraints.

\subsubsection{Control algorithms co-designed with computing resources}
As control systems increasingly rely on data-intensive decision-making, computing resources play an important role in QoS-aware control. With edge/fog computing, the computing load can be distributed to process control tasks closer to where they are needed, hence reducing the dependency on cloud resources~\cite{Tang10354444}. This mechanism reduces latency and allows control systems to act faster, and enhances the real-time responsiveness of ICPS. Autonomous ICPS will require control algorithms that are tightly coupled with computing systems, and also dynamically adjust based on the availability of distributed computing resources. \R{Through 3C co-design, future QoS-aware control systems will achieve higher levels of efficiency, scalability, and adaptability, hence meeting the increasingly complex demands of industrial automation and real-time operations.}

\section{Cyber-Physical Security}
\label{section6_security}

Cyber-physical security focuses on protecting industrial control systems from cyber threats that can disrupt physical processes. In the context of autonomous ICPS, this security aspect becomes even more critical due to the increased connectivity, integration, and convergence of IT/OT. Ensuring the security of these systems involves protecting the integrity, availability, and confidentiality of both cyber and physical components. Cyber-physical security strategies must address threats that can affect everything from data networks to physical machinery, potentially causing significant operational disruptions, safety hazards, and financial losses.

\begin{figure}
  \begin{center}
   \includegraphics[trim={0cm 0cm 0cm 0cm},clip,width=0.8\linewidth]{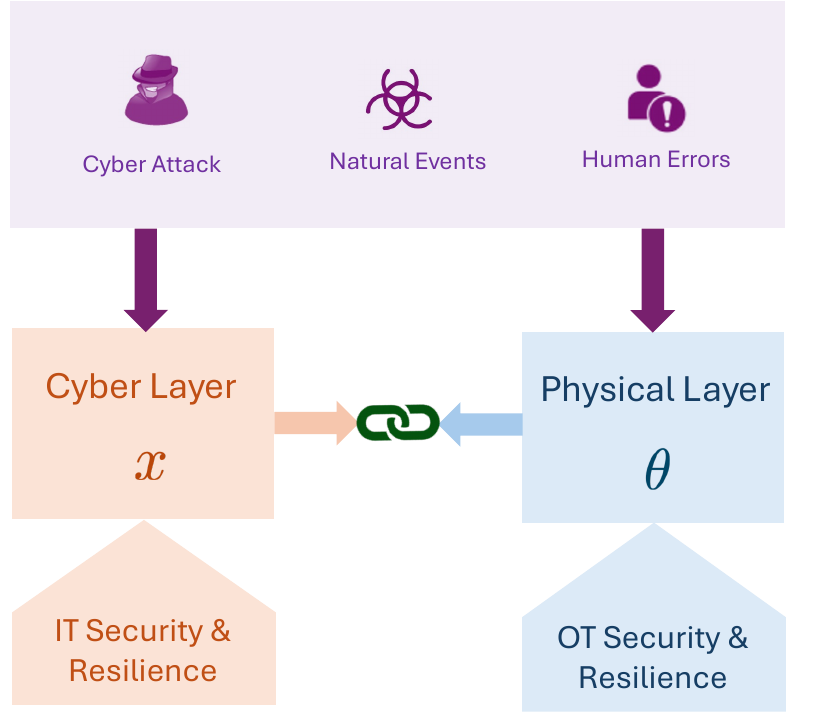}
  \end{center}
 \caption{\R{An industrial control system representing the convergence of its cyber and physical layers. The state of the well-being of the IT system $x$ and the state of the OT system $\theta$ are interdependent. They are exposed to a broad attack surface, including DDoS attacks, ransomware, malware, jamming, MitM attacks, and human vulnerabilities. Security and resilience mechanisms must be  designed at each layer to account for the  cross-layer interactions and inter-dependencies. A co-design approach is essential to improve the cyber-physical security of the IT and OT-convergent ICPS.}}  
\end{figure}

The need for cyber-physical security co-design arises from the inherently intertwined nature of ICPS, which is divided into distinct yet interdependent cyber and physical layers. The \textit{cyber layer} comprises assets such as servers, workstations, sensors, communication networks, and networking components. In contrast, the \textit{physical layer} includes actuators, controllers, and factory plants. This tight coupling requires an integrated security approach that considers the mutual dependencies and interactions between these layers. One example of this tight coupling is in large-scale power networks. When a cyber attack targets the communication network, it could disrupt the data that is being transmitted from the sensors to the control systems. This disruption might lead to incorrect readings or commands being sent to actuators and controllers. As a result, an attacker could potentially cause power outages, equipment damage, or even trigger cascading failures across the network~\cite{li2012securing}.

Historically, many legacy systems in critical infrastructures are isolated from networks, making it difficult for attackers to access sensitive physical system information. However, the integration of ICPS with cloud computing and other modern information technologies has exposed these systems to a wide range of cyber attacks. \R{The adoption of cloud computing in ICPS has brought numerous benefits, such as enhanced scalability, improved data analytics, and cost efficiency. Cloud computing enables real-time monitoring, remote control, and data storage, which are essential to optimize industrial processes and improve operational efficiency.} Despite these advantages, cloud computing also introduces new vulnerabilities and attack vectors~\cite{ali2015security}. Traditional cloud security solutions, which focus primarily on data confidentiality and integrity, are insufficient for ICPS protection due to the critical need for data availability and real-time data for system stability. ICPS operations are highly dependent on timely and accurate flow of information to control and monitor physical processes. \R{Any delay or interruption can have severe operational and safety implications. For instance, latency issues in cloud-based communications can disrupt the synchronization of control systems, leading to potential malfunctions or safety hazards.}

Modern attackers employ sophisticated methods, such as advanced persistent threats (APTs), which can infiltrate systems over extended periods, compromising secret keys and sensitive information before executing further stages of an attack. Additionally, the inter-dependencies between the cyber and physical layers present further challenges. \R{Effective ICPS security requires a holistic approach where both layers are co-designed. The cyber layer must support the physical layer's performance demands, while the physical layer must ensure that the cyber layer's security measures are robust and resilient.} This integrated design improves the overall security and resiliency of ICPS. To address these challenges, co-designed security frameworks that consider the unique requirements of ICPS environments are pertinent. These frameworks should integrate advanced threat detection and response capabilities, leveraging AI/ML and data analytics to identify and mitigate potential threats in real time.

\subsection{Cyber-Physical Security Threats}

\R{Cyber-physical security issues in cloud-enabled industrial automation systems present significant challenges due to the complex and interconnected nature of these environments. Industrial control systems, including SCADA systems and distributed control systems (DCS), are integral components of industrial automation. These systems are increasingly integrated with cloud computing and fog computing architectures to enhance efficiency, scalability, and real-time decision-making. However, this integration also introduces new vulnerabilities and attack vectors. Key threats include Denial-of-Service (DoS) and Distributed Denial of Service (DDoS) attacks, which disrupt system availability by overwhelming resources, and Man-in-the-Middle (MitM) attacks, where attackers intercept and manipulate communications, leading to unauthorized access and compromised system responses. In addition, APTs pose complex, prolonged risks that exploit vulnerabilities over time to steal data or disrupt operations. Ransomware attacks target ICPS' dependency on data availability, encrypting critical information, and demanding payments for decryption, while supply chain attacks compromise systems indirectly by exploiting vulnerabilities in the production or delivery chain of hardware and software. Finally, physical security breaches allow attackers to manipulate or disrupt systems through direct access to components such as servers and control hardware, often complementing cyber attacks to amplify their impact.}

\R{Addressing these threats requires a comprehensive, integrated security approach. This includes robust defenses such as real-time monitoring, advanced detection mechanisms, and co-design strategies that combine physical and cyber security measures. Innovations such as blockchain for supply chain integrity, zero-trust architectures, and multi-layered models are being explored to enhance resilience. A unified understanding of the interconnected nature of these threats is critical for developing strategies that mitigate risks while maintaining system performance, ensuring the integrity of ICPS against an evolving and complex threat landscape.}

\subsection{State-of-the-art of Cyber-Physical Security Co-design}

\R{The co-design of cyber-physical security solutions requires impact-aware strategies that account for how security measures affect overall system performance, such as communication delays or signal-to-noise ratio (SNR). For example, the work in~\cite{zhu2013networked} explores impact-aware power control mechanisms to optimize network control systems alongside wireless communication networks. This approach improves both system efficiency and resilience against adversarial behaviors, highlighting the importance of designing security measures that balance robust protection with minimal impact on overall system performance.}

\R{In addition to ensuring that cyber designs are impact aware, it is equally important to design physical control systems taking into account the performance limitations imposed by the cyber layer. The effective integration of these layers is crucial to optimizing the overall performance of the system. For example, communication-aware control design is a key approach for developing networked control systems. This method involves designing control strategies that account for communication constraints such as delays and packet drop rates, which can significantly impact control performance. Several studies have contributed to this field, including~\cite{yuksel2013stochastic,moon2014control,zhang2015survey}, which explore optimal control performance by integrating communication parameters into the control design.}
 

\subsection{Design for Resilience}

\R{In cloud-enabled industrial automation systems, traditional security measures must evolve to focus on resilience, as perfect cyber-physical security is unattainable and vulnerabilities are inevitable. Resilience-oriented design emphasizes building systems that can withstand and recover from cyber-attacks with minimal operational disruption~\cite{zhu2024disentangling}. For example, recent research highlights mechanisms to quantify and mitigate risks in cloud-enabled networked control systems (CE-NCS)~\cite{xu2015secure}, which often outsource heavy computations to cloud servers. These systems face challenges in ensuring trustworthiness in their cyber-physical connections and cloud dependencies, necessitating robust strategies to maintain operational integrity.}

\R{Redundancy is a key resilience strategy that involves backup systems and duplicate network paths to ensure continuity even when parts of the system are compromised. Studies on IoT networks and the Internet of Battlefield Things (IoBT) have demonstrated the value of adaptive network designs that can reconfigure dynamically in response to disruptions. These designs improve robustness against threats such as DoS and jamming by incorporating optimal strategies that adapt to changing network topologies, ensure continuous operations, and safeguard critical processes.}

Self-healing adaptive systems further bolster resilience by detecting, isolating, and repairing damage from cyber-attacks in real time. These systems use automated response mechanisms to maintain functionality without human intervention, adapting to evolving threats and minimizing disruption. By integrating these resilience-by-design frameworks with traditional cybersecurity measures, organizations can create robust, adaptive defenses that maintain functionality and recovery capabilities in the face of persistent and evolving cyber threats. This holistic approach is essential for the security of complex industrial environments.

\subsection{Future Directions}

Recent rapid advances in AI and 5G communication networks have introduced a wide range of promising research directions in cyber-physical security and system resilience. As these technologies evolve, they offer both unprecedented opportunities and new challenges for securing complex and interconnected systems. The integration of 5G technology into ICPS transforms their operational capabilities, allowing for more efficient, real-time communication and data processing. However, this also expands their attack surface, making them more vulnerable to cyber threats.

\subsubsection{Research directions enabled by 5G}
The enhanced connectivity offered by 5G means that ICPS components, such as sensors, controllers, and actuators, are more interconnected than ever before. While this promotes greater automation and efficiency, it also introduces new entry points for attackers. As more devices and systems communicate with the cloud and each other, the number of vulnerable nodes increases, creating more opportunities for attackers to exploit weak points in the network. The larger and more complex the system becomes, the greater the security challenges.

As 5G enables real-time data processing in the cloud, securing cloud infrastructure becomes more critical. Cloud platforms are susceptible to a variety of threats, including data breaches, misconfigurations, insider threats, and sophisticated attacks such as hyper-jacking, where attackers exploit vulnerabilities in virtual machines. If attackers compromise the cloud infrastructure, they can gain unauthorized access to critical ICPS data or manipulate industrial processes remotely, posing significant risks to both cybersecurity and physical safety.

The increased reliance of 5G networks on software and virtualization also makes them more susceptible to MitM attacks. In a cloud-enabled ICPS, attackers could intercept communications between cloud services and industrial devices, alter data, or inject malicious commands, potentially leading to disruptions in critical operations or creating safety hazards in physical processes. Given these risks, there is an urgent need for research and solutions to protect 5G-enabled ICPS environments. One key strategy is network segmentation~\cite{chahbar2020comprehensive}, which can isolate critical industrial systems from less secure or public networks. Micro-segmentation techniques further enhance this by containing potential attacks and preventing them from spreading across the ICPS. In addition, zero-trust security models~\cite{stafford2020zero} are essential to ensure that no device, user, or application is trusted by default, even if it is within the network perimeter. Continuous verification, strict access controls, and multi-factor authentication can significantly reduce the risk of unauthorized access to critical systems.

AI-driven security solutions are also pivotal in securing 5G-enabled ICPS. By monitoring network traffic, detecting anomalies, and predicting potential threats in real-time, AI can help identify and mitigate cyberattacks before they cause significant damage. In this evolving landscape, AI-based tools play an increasingly crucial role in strengthening the security and resilience of industrial control systems.

\subsubsection{Research directions driven by AI}
Recent advances and excitement in foundation models, such as LLMs and deep learning architectures, further increase the power of prediction and automated responses in handling security concerns~\cite{qiu2023promise}. These models can process and understand vast amounts of data, learning intricate patterns that simpler models might miss. By integrating foundation models into security systems, organizations can achieve more accurate threat detection and more sophisticated response strategies. For example, foundation models can improve the understanding of complex attack vectors, enabling more effective defense mechanisms that adapt to evolving threats. The work presented in~\cite{qiu2023promise} explores an innovative approach to the design of foundation models for the management of cloud systems, focusing particularly on meta-learning techniques.


Recent work in~\cite{li2024symbiotic} has introduced an innovative approach that combines integrated game-theoretic models with foundation models to improve data analytics for cybersecurity. This research aims to develop descriptive and prescriptive analytics to design proactive deception mechanisms, specifically targeting the mitigation of APT attacks. By leveraging game theory, the study provides a strategic framework for understanding the interactions between attackers and defenders, while the use of foundation models enhances the adaptability and effectiveness of these strategies. The proposed approach not only identifies potential attack vectors but also prescribes optimal countermeasures that can be implemented in real-time, significantly reducing the impact of APT attacks on critical systems. This integration of advanced modeling techniques is a promising direction in the development of resilient and proactive cybersecurity solutions.

The work presented in~\cite{brohan2023can} improves the practical application of LLMs in real-world robotic tasks by proposing a novel approach that integrates LLMs with pre-trained robotic skills. This integration allows the language model to generate actions in natural language that are not only contextually appropriate but also feasible within the robot's operational environment. Essentially, the robot acts as the ``hands and eyes" of the language model, translating abstract/high-level instructions into concrete/executable tasks. This method effectively marries the semantic understanding of LLMs with the grounded, environment-specific capabilities of robotic systems, enabling robots to carry out complex, long-horizon tasks guided by natural language instructions. The success of this approach has been demonstrated through rigorous evaluations of real-world robotic tasks, demonstrating its potential to significantly advance the field of human-robot interaction.

The use of AI in protecting ICPS offers tremendous potential to improve security and resilience, but also introduces its own set of vulnerabilities. AI can improve threat detection, automate responses, and provide predictive analytics to prevent security incidents in ICPS. However, the very algorithms that power AI are susceptible to attacks, such as data poisoning, where attackers introduce malicious data into the training process, causing the model to make incorrect predictions or classifications. This can have serious consequences in an industrial environment, where incorrect decisions could disrupt critical operations, cause safety hazards, and even cause damage to the physical infrastructure. Manipulation attacks on AI models, particularly LLMs, present another significant risk. Adversaries can exploit LLMs by manipulating the inputs to achieve desired but incorrect outputs, potentially misleading decision-makers, or causing automated systems to behave in unintended ways. For example, LLMs could be used to help manage complex operations or provide recommendations for system adjustments, but if an attacker manipulates the model, the results could lead to operational inefficiencies or even critical system failures.

It is crucial to be aware of these AI vulnerabilities while simultaneously developing secure and resilient solutions for ICPS. AI's increasing role in ICPS security demands a comprehensive approach that not only incorporates AI-based defenses but also actively addresses the specific threats posed to AI itself. Robust defenses such as adversarial training, data integrity checks, and secure model development processes must be used to mitigate the risks of data poisoning and model manipulation. The holistic integration of AI, cyber, and physical ICPS stacks creates additional layers of complexity. The interplay between these elements introduces new challenges in the future design of secure systems. For example, AI may monitor and control cyber elements, such as network traffic, while simultaneously interfacing with physical components, such as sensors and actuators. Ensuring that each layer, i.e., AI algorithms, cybersecurity protocols, and physical controls, works harmoniously and securely together requires a multidisciplinary cyber-secure system design approach.

\section{Concluding Remarks}
\label{conclusions}

\R{Recent research and development efforts in advancing network-centric industrial automation reference architectures such as Cloud-Fog Automation have prompted us to explore its broader and more futuristic applicability. In this paper, we have presented a comprehensive overview of the foundations of Cloud-Fog Automation as the new paradigm for realizing the vision of a fully autonomous ICPS, as well as highlighting the crucial role of communication, computing, and control co-design, i.e., ``3C co-design". We first introduced a Cloud-Fog Automation prototype which focused on four representative industrial use cases. We then outlined the vision and design requirements of autonomous ICPS, presented research gaps and discussed how 3C co-design is poised to close the gap between the new-generation digital infrastructure and future autonomous ICPS. We subsequently examined the state-of-the-art in goal-oriented communication, virtualization-empowered computing, QoS-aware control; and identify key opportunities and future research directions to foster innovation and bridge the gap between theory and practical applications in these inter-related domains. Finally, we emphasized the critical importance of cyber-physical security as a cross-cutting concern across the communication, computing, and control domains.}

\R{There are tremendous opportunities, open research challenges and future directions in delivering an integrated and holistic 3C co-design for autonomous ICPS.} It is our sincere hope that this paper serves as both a roadmap and a call-for-action for the community to steer research and development efforts in the field, with a strong emphasis on the pivotal role of 3C co-design in driving Cloud-Fog Automation as the foundational paradigm for future autonomous ICPS.

\ifCLASSOPTIONcaptionsoff
  \newpage
\fi


\bibliographystyle{IEEEtran}
\bibliography{IEEEabrv,
References/jsac_full_references
}

\begin{IEEEbiography}[{\includegraphics*[width=1in,height=1.25in,clip,keepaspectratio]{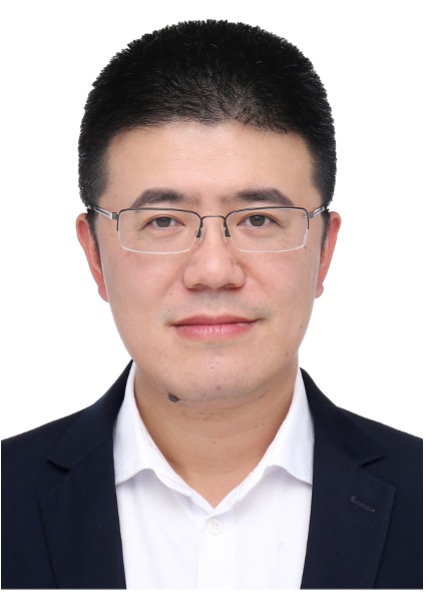}}]{Jiong Jin} (Member, IEEE) received the B.E. degree with First Class Honours in Computer Engineering from Nanyang Technological University, Singapore, in 2006, and the Ph.D. degree in Electrical and Electronic Engineering from the University of Melbourne, Australia, in 2011. He is currently a full Professor in the School of Engineering, Swinburne University of Technology, Melbourne, Australia. His research interests include network design and optimization, edge computing and intelligence, robotics and automation, and cyber-physical systems and Internet of Things as well as their applications in smart manufacturing, smart transportation and smart cities. He was recognized as an Honourable Mention in the AI 2000 Most Influential Scholars List in IoT (2021 and 2022). He is currently an Associate Editor of IEEE Transactions on Industrial Informatics and IEEE Transactions on Network Science and Engineering.
\end{IEEEbiography}

\vskip -2\baselineskip plus -1fil

\begin{IEEEbiography}[{\includegraphics*[width=1in,height=1.25in,clip,keepaspectratio]{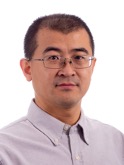}}]{Zhibo Pang} (Senior Member, IEEE) received MBA in Innovation and Growth from University of Turku in 2012 and PhD in Electronic and Computer Systems from KTH Royal Institute of Technology in 2013. He is Senior Principal Scientist at ABB Corporate Research Sweden and Adjunct Professor at KTH Royal Institute of Technology and was Adjunct Professor at University of Sydney. He is a Member of IEEE IES Industry Activities Committee, Steering Committee Member of IEEE IoT Technical Community, Chair of IEEE Technical Committee on Cloud and Wireless Systems for Industrial Applications, and Co-Chair of IEEE TC on Industrial Informatics. He is Associate Editor of IEEE TII, IEEE JBHI, IEEE TCE, IEEE TSUSC, IEEE JESTIE, and IEEE IoTM. He was awarded ``Inventor of the Year Award" by ABB Corporate Research Sweden, three times in 2016, 2018, and 2021 respectively. He works on embodied intelligence, robotics, control, computing, communication, and electronics for Industry 4.0 and Healthcare 4.0. He has many productized research results and 23 granted patents in US, Europe, or Japan.
\end{IEEEbiography}

\vskip -2\baselineskip plus -1fil

\begin{IEEEbiography}[{\includegraphics*[width=1in,height=1.25in,clip,keepaspectratio]{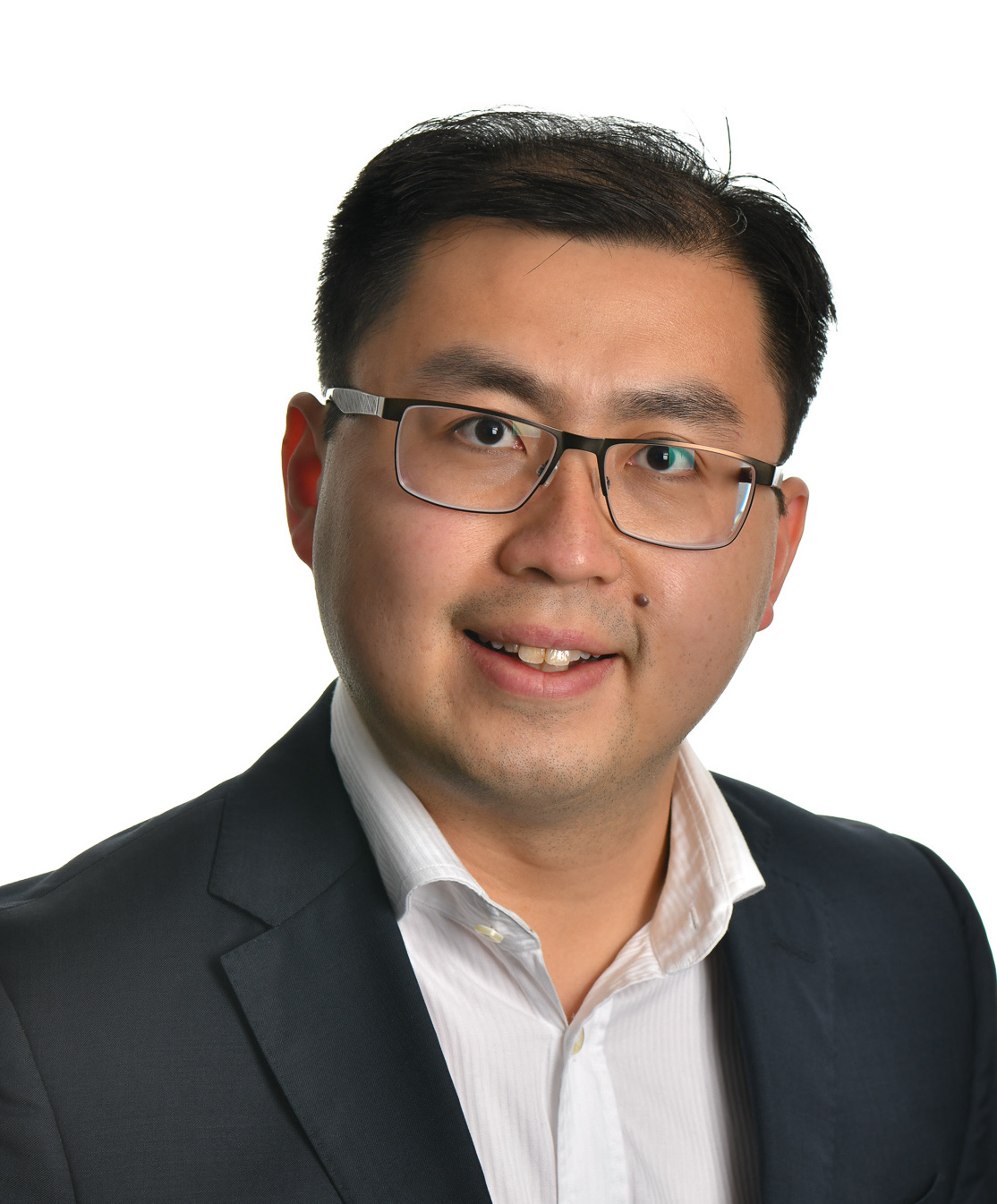}}]{Jonathan Kua}(Member, IEEE) received the B.Eng. (First Class Hons.) degree in telecommunications and network engineering and the Ph.D. degree in telecommunications engineering from Swinburne University of Technology, Australia, in 2014 and 2019, respectively. He is currently Senior Lecturer in Internet of Things within the School of Information Technology, Deakin University, Australia. His research interests include low-latency data transport protocols, adaptive streaming, content delivery, Internet of Things and industrial cyber-physical systems, with an overarching focus on improving their network performance and quality of service. He was the recipient of the Netflix Ph.D. scholarship award (2015–2019) and the second runner-up of the DASH-IF Best Ph.D. Dissertation Award on ``Algorithms and Protocols for Adaptive Content Delivery over the Internet" (2019).
\end{IEEEbiography}

\vskip -2\baselineskip plus -1fil

\begin{IEEEbiography}[{\includegraphics*[width=1in,height=1.25in,clip,keepaspectratio]{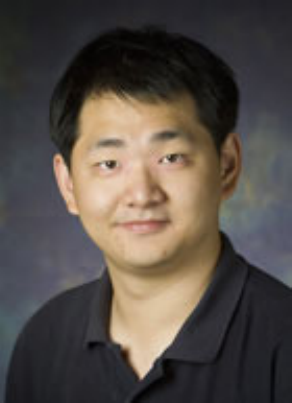}}]{Quanyan Zhu} (Senior Member, IEEE) is an Associate Professor in the Department of Electrical and Computer Engineering at New York University (NYU). He is an affiliated faculty member with the Center for Urban Science and Progress (CUSP) and the Center for Cyber Security (CCS) at NYU. His research interests include cyber and physical systems, multi-agent systems, and cybersecurity and resilience. He serves as an Associate Editor for IEEE Transactions on Aerospace and Electronic Systems and IEEE Transactions on Network Science and Engineering. He currently serves as the technical committee chair on security and privacy for the IEEE Control Systems Society.
\end{IEEEbiography}

\vskip -2\baselineskip plus -1fil

\begin{IEEEbiography}[{\includegraphics*[width=1in,height=1.25in,clip,keepaspectratio]{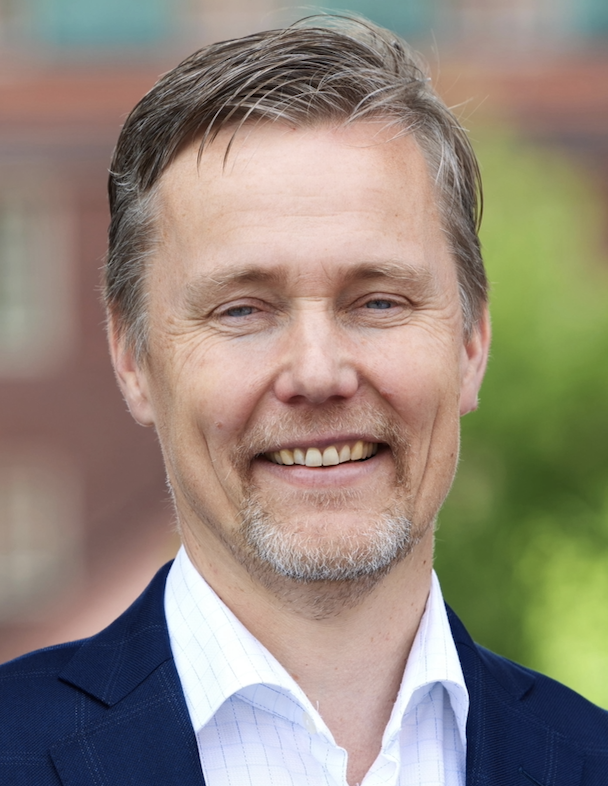}}]{Karl H. Johansson} (Fellow, IEEE) is Swedish Research Council Distinguished Professor in Electrical Engineering and Computer Science at KTH Royal Institute of Technology in Sweden and Founding Director of Digital Futures. He earned his MSc degree in Electrical Engineering and PhD in Automatic Control from Lund University. He has held visiting positions at UC Berkeley, Caltech, NTU and other prestigious institutions. His research interests focus on networked control systems and cyber-physical systems with applications in transportation, energy, and automation networks. For his scientific contributions, he has received numerous best paper awards and various other distinctions from IEEE, IFAC, and other organizations. He has been awarded Distinguished Professor by the Swedish Research Council, Wallenberg Scholar by the Knut and Alice Wallenberg Foundation, Future Research Leader by the Swedish Foundation for Strategic Research. He has also received the triennial IFAC Young Author Prize, IEEE CSS Distinguished Lecturer, IFAC Outstanding Service Award, and IEEE CSS Hendrik W. Bode Lecture Prize. His extensive service to the academic community includes being President of the European Control Association, IEEE CSS Vice President Diversity, Outreach \& Development, and Member of IEEE CSS Board of Governors and IFAC Council. He has served on the editorial boards of Automatica, IEEE TAC, IEEE TCNS and many other journals. He has also been a member of the Swedish Scientific Council for Natural Sciences and Engineering Sciences. He is Fellow of both the IEEE and the Royal Swedish Academy of Engineering Sciences.
\end{IEEEbiography}

\vskip -2\baselineskip plus -1fil

\begin{IEEEbiography}[{\includegraphics*[width=1in,height=1.25in,clip,keepaspectratio]{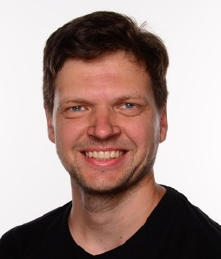}}]{Nikolaj Marchenko} (Member, IEEE) is a Researcher Engineer at Robert Bosch GmbH in Stuttgart, Germany, where he works on the topics of the future wireless networked systems in industrial automation and automotive domains, with particular focus on joint optimization and co-design. Nikolaj received his PhD in Information Technology at Klagenfurt University (Austria) in 2013, and his Diploma Degree in Computer Engineering from RWTH Aachen University (Germany) in 2007.
\end{IEEEbiography}


\vskip -2\baselineskip plus -1fil

\begin{IEEEbiography}[{\includegraphics*[width=1in,height=1.25in,clip,keepaspectratio]{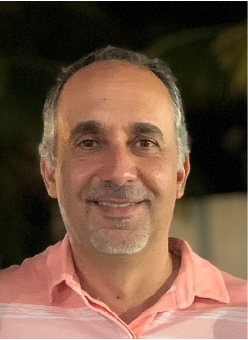}}]{Dave Cavalcanti} (Senior Member, IEEE) is a Principal Engineer at Intel Corporation with extensive experience in distributed networked systems, wireless connectivity, industry standards and ecosystems. Dave joined Intel in 2015, and he also serves as President of the Avnu Alliance, an industry forum driving standards and certification programs to enable deterministic performance and Time Sensitive Networking (TSN) across Ethernet/Wi-Fi/5G connectivity. He has contributed to several generations of the IEEE 802.11/Wi-Fi standards. He received a PhD in Computer Science and Engineering in 2006 from the University of Cincinnati. He has published over 50 peer-reviewed papers and holds 99 granted patents. He is a Senior member of the IEEE and holds several leadership positions in IEEE conferences and publications. He is the recipient of best paper awards at IEEE WFC'22, WFCS'21, IEEE INDIN'21, and the best demo award at IEEE INFOCOM 2018.
\end{IEEEbiography}

\end{document}